\documentclass[12pt]{article}
\usepackage[ hmargin=3 cm,vmargin=3cm]{geometry}

 \usepackage{graphicx}
 \usepackage{amsmath}
 \usepackage{amssymb}
  \usepackage{enumerate}
\usepackage{cite}

\usepackage{float}
\usepackage{tikz}
\usetikzlibrary{decorations.pathmorphing}
\usepackage[symbol]{footmisc}

\newcommand{\be}{\begin{eqnarray}}
\newcommand{\ee}{\end{eqnarray}}
 \begin{document} \title{Multi-Time Measurements in Hawking Radiation: Information at Higher-Order Correlations  } \author {Charis Anastopoulos\footnote{anastop@physics.upatras.gr}  \;   and   \; Ntina Savvidou\footnote{ksavvidou@upatras.gr}\\
 {\small Department of Physics, University of Patras, 26500 Greece} }

\maketitle

\begin{abstract}
It is believed that no information can be stored in    Hawking radiation, because  correlations between quanta of different field modes vanish.   However, such correlations have been defined only with reference to a single moment of time. In this article, we develop a method for the evaluation of {\em multi-time} correlations. We find that these correlations are highly non-trivial: for a scalar field in the Schwarzschild black hole, multi-time correlations have
 an explicit dependence on angular variables and on the scattering history of Hawking quanta. This result leads us to the  conjecture that   some pre-collapse information can be stored in multi-time  correlations after backreaction effects have been  incorporated in the physical description.
\end{abstract}

\section{Introduction}
Statistical correlations between different observables are well defined in any probabilistic theory. In quantum theory, correlations between different components of  a multi-partite system, or different locations in an extended  quantum system (e.g., a quantum field) can be experimentally determined, and reveal substantial information about the system.

In the context of black hole thermodynamics, it is widely believed that  correlations between Hawking
 quanta emitted by a black hole carry no information. The reason is that the reduced state of the quantum field  far from the horizon  is asymptotically Gibbsian at the Hawking temperature, modulo the grey-body factor. This is the content of the Hawking-Wald (HW) theorem \cite{Hawk1, Wald1}, described in detail in Sec. 2. Hence, the values of all physical observables are distributed thermally at late times. This means that there is no correlation  between quanta of different field modes. However,
the statements above apply only to single-time correlations of the emitted radiation, i.e., to sets of measurements that are carried out at a single  instant of time.

In this article, we show that there are no such constraints for {\em multi-time} measurements, i.e., sets of measurements carried out at different instants of time.  We present a general method for the evaluation of multi-time correlations, and we employ it for a scalar field in a Schwarzschild black hole.
In particular, we show that multi-time correlations can be non-thermal, and that they have a   complex form that allows them to store non-trivial information. Our results demonstrate that  any discussion of the informational balance  in the process of black hole formation and evaporation must take the existence of multi-time correlations into account.
  We argue that multi-time correlations can, in principle, preserve some memory of characteristics of the system prior to collapse. This conjecture will be tested by a successful treatment of backreaction.

Our analysis is based on probability formulas for multi-time measurements on a quantum field. We derive these formulas  by modeling the measuring apparatuses
 as
 generalized Unruh-Dewitt (UdW) detectors  \cite{Unruh76, Dewitt, HLL12}. An UdW detector is a  {\em pointlike} quantum system that moves along a specific trajectory $X(\tau)$ with proper time $\tau$, and it is coupled to a quantum  field through a term   $\hat{O}(X(\tau))$, where $\hat{O}(X)$ is a local composite operator for the field. In our model, the detector records quanta of a quantum field, but also the proper time at which the detection took place. Hence, by considering multiple UdW detectors at different trajectories, we can describe multi-time measurements of quantum field.

The UdW detector model involves a coupling of particle and field degrees of freedom  that cannot be justified from first principles. However, to leading order in perturbation theory and for vanishing detector size, the model leads to the same results with a more rigorous quantum measurement formalism that involves genuine QFT couplings between measured system and apparatus  \cite{AnSav12, AnSav13, AnSav17, AnSav19}. The use of the UdW detector model allows us to present
   an elementary and self-contained  derivation of correlations in  Hawking radiation.

As a first application of our model, we study multi-time correlations  in the Unruh effect. To this end, we consider
 two accelerated detectors in Minkowski spacetime. This system was first studied in Ref.  \cite{AnSav11}, where it was shown that the correlations recorded by the detectors are not thermal, thus, providing a motivation for the present work. We give a simpler proof of this result, discuss its implications, and analyse the structure of the correlation function in different regimes.

Then, we study Hawking radiation from an eternal black hole with a quantum field in the Unruh vacuum. The latter  simulates the late time behavior of a quantum field state   in a collapsing black hole spacetime   \cite{Unruh76}. We derive a general formula for the two-time coincidence function, that quantifies the quantum correlations between two  detection events at different spacetime points.  These correlations are in general non-causal, in the same sense that Bell-type correlations are non-causal.
We find a complex dependence of those correlations on the location of the detectors,  on energy and on the scattering history of the Hawking quanta in the Schwarzschild potential. In particular, the effect of the Schwarzschild potential cannot be incorporated in a single function of energy as in the case of single-time measurements.
This suggests that multi-time correlations could store significant pre-collapse information, once the backreaction of the field to the spacetime geometry is taken into account.



The structure of this paper is the following. In Sec. 2,  we provide a detailed analysis of the HW theorem showing that it does not constrain multi-time correlations. In Sec. 3, we discuss  multi-time measurements in QFT, and we  derive the probability formulas for generalized UdW detectors. In Sec. 4, we revisit the issue of multi-time correlations  in the Unruh effect. In Sec. 5, we evaluate temporal correlations in the Hawking radiation of eternal  black holes. In Sec. 6,  we present the physical interpretation of our results.

\section{The HW theorem and its limitations}

  In the original analysis of black hole radiation, Hawking proved that particle numbers at the future null infinity $\mathcal{I}^+$ are characterized by a Planckian spectrum \cite{Hawk1}. Subsequently, Wald showed that {\em all} observables at $\mathcal{I}^+$ behave thermally \cite{Wald1}. We   refer to the latter result as the Hawking-Wald (HW) theorem. It is based on a scattering-matrix approach to quantum field theory (QFT) in curved spacetime. It implies that there is no correlation between different field modes of the emitted radiation \cite{Park75, Hawk76}.

  The HW theorem is commonly cited as a proof  that no information can be stored in the correlations of Hawking radiation. However, the theorem refers only to asymptotic single-time properties of the quantum field, and it makes no  statement about multi-time measurements at late times. In this section, we present a  detailed analysis of the theorem that makes our point explicit.

This is an independent section: its definitions and results are not used elsewhere in this paper. The reader interested in the description of multi-time measurements may skip directly to section 3.

\begin{figure}
    \centering
    \begin{tabular}{ll}
 \includegraphics[width=0.34\textwidth]{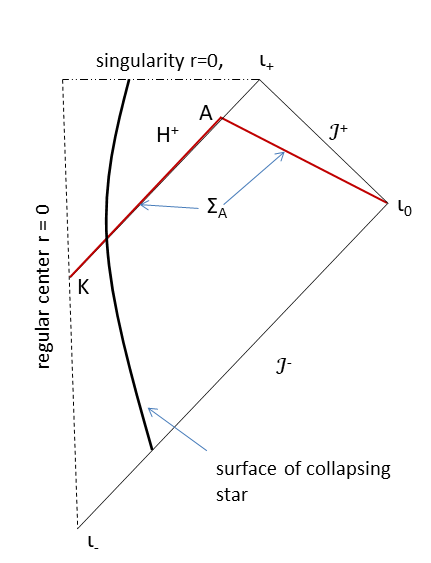} \hspace{1cm}
 \includegraphics[width=0.34\textwidth]{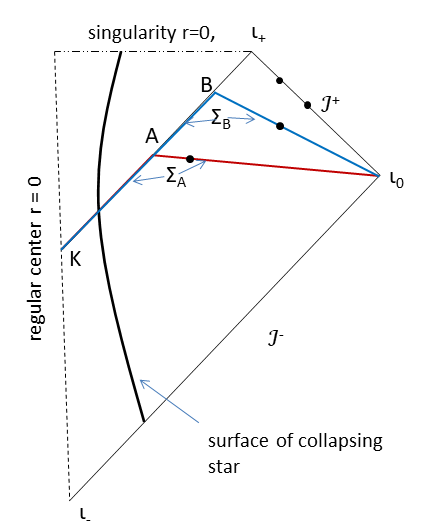}
 \end{tabular}
    \caption{Left: Penrose diagram of a collapsing black hole. Right: Sequential events of particle detection (black dots) in a collapsing black hole spacetime.}
    \label{fig:foobar}
\end{figure}

\medskip

\subsection{Preliminaries}

Let $(M, g)$ be an asymptotically flat spacetime  that describes the collapse of a star leading to the formation of a black hole with future event horizon ${\cal H}^+$. Consider a free quantum scalar field $\hat{\Phi}(X)$ on $M$. To construct the  Hilbert space ${\cal F}$  of states for this field, one first identifies the real vector space $V$ of solutions to Klein-Gordon's (KG) equation with the KG inner product. Then, one complexifies $V$ in order to construct a complex Hilbert space $V_{\pmb C}$ \cite{QFTCSWald}. The Hilbert space ${\cal F}$ for the field degrees of freedom  is the exponential Hilbert space $e^{V_{\pmb C}}$, i.e., the bosonic Fock space associated to $V_{\pmb C}$
\begin{eqnarray}
{\cal F} = e^{V_{\pmb C}} := {\pmb C}\oplus V_{\pmb C} \oplus (V_{\pmb C} \otimes V_{\pmb C})_S \oplus  (V_{\pmb C} \otimes V_{\pmb C} \otimes V_{\pmb C} )_S \oplus \ldots,
\end{eqnarray}
where the index $S$ refers to symmetrization. To complexify $V$, one chooses a subset of complex-valued solutions $u_a(X)$  to the KG equation   to define an orthonormal basis on $V_{\pmb C}$.

In presence of a time-like Killing vector $\partial/\partial t$, the functions $u_a(X)$   are positive-frequency with respect to $t$,
 \begin{eqnarray}
 i \frac{\partial}{\partial t} u_a = \omega_a u_a, \label{posfrq}
 \end{eqnarray}
 for $\omega_a > 0$.

Once the family of solutions $u_a(X)$ has been chosen, the field operator is expressed as
\begin{eqnarray}
\hat{\phi}(X) = \sum_a \left[ \hat{a}_a u_a(X) + \hat{a}^{\dagger}_a u^*_a(X) \right],
\end{eqnarray}
in terms of the creation and annihilation operators on  the Fock space ${\cal F}$.

\subsection{The field as a bipartite system}
 Let $W$ be a closed linear subspace of $V_{\pmb C}$, and $W^{\bot}$ its complement. Then, the field Hilbert space splits as a tensor product \cite{Klau70}
 \begin{eqnarray}
 {\cal F}  = e^{W\oplus W^{\bot}} = e^{W} \otimes e^{W^{\bot}}. \label{split}
 \end{eqnarray}
Of particular relevance are tensor products of the form Eq. (\ref{split}) that are generated by partitioning a Cauchy surface.
Let $\Sigma$  be a Cauchy surface on $M$,  and  $C_1$, $C_2$ subsets of $\Sigma$, such that $C_1 \cap C_2 = \emptyset$ and $C_1 \cup C_2 = \Sigma$. We define a subspace $W_{C_1} $ of $V_{\pmb C}$ that consists of all functions $f(X) = \sum_a c_a u_a(X)$, such that $f(X) = 0$, for $X \in C_2$; $W_{C_2} = W_{C_1}^{\bot}$ is spanned by all functions $g(X)$ that vanish for $X \in C_1$.
The Hilbert space ${\cal H}_{C_i} := e^{W_{C_i}}$ describes field states localized in $C_i$, $i = 1, 2$. Hence, Eq. (\ref{split})  can be expressed as
\begin{eqnarray}
{\cal F} = {\cal H}_{C_1} \otimes {\cal H}_{C_2}. \label{split2}
\end{eqnarray}

The field Hilbert space splits like  a bipartite system. This split is essential for the derivation of the HW theorem, and, consequently, of our discussion of its limitations.  However, the interpretation of Eq. (\ref{split2}) as a physical bipartite system requires the statistical independence of measurement outcomes and of state preparations in $C_1$ and $C_2$. The relevant condition is the so called "split property" of Borchers and Bucholz \cite{BoBu}---see, also Ref. \cite{Few} for a  review. The split property requires that the two regions do not touch at their boundaries, which is not the case here. This implies that
caution should be exercised when using notions that treat the field as a bipartite system, like, for example, entanglement entropy or theorems about the properties of entanglement.

 We choose a basis $f_i(X)$ of solutions in $W$, such that  $f_i(X) = 0$ for $X \in C_2$, and a basis $g_j (X)$of solutions in $W^{\bot}$, such that  $g_j(X) = 0$ for $X \in C_1$. Then, the field operator can be written as
 \begin{eqnarray}
\hat{\phi}(X) = \sum_i \left[ \hat{a}^{(1)}_i f_i(X) + \hat{a}^{(1)\dagger}_i f^*_i(X) \right] + \sum_j \left[ \hat{a}^{(2)}_j g_j(X) + \hat{a}^{(2)\dagger}_j g^*_j(X) \right]
\end{eqnarray}
 where $\hat{a}^{(1)}$ and   $\hat{a}^{(2)}$ are annihilation operators restricted to the subspace $W$ and $W^{\bot}$ respectively.

Any state $|\Psi\rangle \in {\cal F}$ can be expressed as a linear combination $\sum_{A, B} \lambda_{AB} |A\rangle_{1}\otimes |B\rangle_{2}$, where $|A\rangle_1 $ and $|B\rangle_2 $ define orthonormal sets in $e^{W}$ and $e^{W^{\bot}}$, respectively. Any vector $|A\rangle_1$ can be constructed from the consecutive  action of creation operators $\hat{a}^{(1)\dagger}_i$ on a reference state $|0 \rangle_1$, and any vector $|B\rangle_2$ can be constructed from the consecutive  action of creation operators $\hat{a}^{(2)\dagger}_j$ on a reference state $|0 \rangle_2$.

Consider now a single-time measurement localized in the region $C_1$.  Suppose that the field interacts with a measuring apparatus through a composite operator
  $\hat{O}(X)$ that is a local functional of the field $\hat{\phi}(X)$. The interaction  depends only on the operators   $\hat{a}_i^{(1)}$ and $\hat{a}^{(1)\dagger}_i$; hence, it affects only the vectors $|A\rangle_{1}$. Therefore, all information about such measurements is contained in the reduced density matrix $\hat{\rho}_1$ on ${\cal H}_{C_1}$,
\begin{eqnarray}
{}_1\langle A|\hat{\rho}_1|B\rangle_1 := \sum_C \lambda_{AC}\lambda^*_{BC}. \label{rho1red}
\end{eqnarray}

In a collapsing black hole spacetime, we consider the Cauchy surface $\Sigma_A := (KA)\cup(A \iota_0)$ of Fig. 1. Let $C_1$ be its segment outside the black hole. After the end of the collapse, there   is a
  time-like Killing vector $\frac{\partial}{\partial t}$ outside the black hole. Hence,  the modes   $f_i(X)$ can be chosen to be positive-frequency with respect to $\frac{\partial}{\partial t}$; the corresponding frequencies $\omega_i$ are then interpreted as single-particle energies. Therefore, we can choose the
basis $|A\rangle_1$ to be eigenstates of the particle number operators, i.e., the elements of the basis are labeled by   a sequence of particle numbers
 $\{n_i\}$ for each mode $i$.

 The key point is that the Hilbert space split above is relevant only for a single moment of time, i.e., a specific Cauchy surface $\Sigma$. Consider a different Cauchy surface $\Sigma'$, with a different split $C_1'$ and $C_2'$.  The functions $g_j(X)$ that vanish in $C_1$ do not, in general, vanish in $C_1'$\footnote[4]{The set of solutions to the KG equation that vanishes on both $C_1$ and $C_1'$ is   of measure zero in $V$.}. It follows that the field operator $\hat{\phi}(X)$ for $X \in C_1'$ does not vanish on $C_2$. A localized measurement at $X \in C_1'$ involves also the operators $\hat{a}^{(2)}_j$ and $\hat{a}^{(2)\dagger}_j$. The probabilities for  two-time measurement localized at $X \in C_1$ and $X' \in C_1'$ cannot be expressed solely in terms of the reduced density matrix, either of the ${\cal H}_{C_1} \otimes {\cal H}_{C_2}$ or of the ${\cal H}_{C'_1} \otimes {\cal H}_{C'_2}$ partition\footnote[3]{Since the Fock space split is time-dependent,  entanglement between modes outside the black hole and on the horizon is also time-dependent. Common statements about this entanglement  refer to its asymptotic value. It is far from obvious that this asymptotic expression remains relevant after the inclusion of backreaction. The effects of backreaction are not asymptotic, for example, the change in the black hole mass is manifested at finite times. However, the Hilbert space split (\ref{split2})  is   not unique at finite times. Any choice of $C_1$ and $C_2$, such that $C_1 \rightarrow  \mathcal{I}^+$ and $ C_2 \rightarrow {\cal H}^+$ leads to the same asymptotic value of entanglement. In our opinion, this is an indication that entanglement may not be the most appropriate measure of the correlations in Hawking radiation, especially in relation to black hole evaporation. One should look for a measure that incorporates information about multi-time correlations and it is uniquely defined at finite times.
  }.

 The HW theorem  involves a split of the form Eq. (\ref{split}) in relation to the Cauchy surface $\Sigma_{\infty} = \mathcal{I}^+ \cup {\cal H}^+$, and it demonstrates that   the reduced density matrix at $\mathcal{I}^+$ is Gibbsian,
 \begin{eqnarray}
 \hat{\rho}_1 = \frac{1}{Z} \sum_{\{n_i\}} e^{-\sum_i n_i \omega_i/T_H} |\{n_i\}\rangle \langle \{n_i\}| \label{hawkrad}
 \end{eqnarray}
where $T_H = (8\pi M)^{-1}$ is the Hawking temperature, and $Z = \sum_{\{n_i\}} e^{-\sum_i n_i \omega_i/T_H}$ is the partition function. The thermal density matrix (\ref{hawkrad}) is approximate as it ignores transient effects, i.e., particles created during collapse. It also assumes unit transmission probability for all field modes under consideration, i.e.,  that all "emitted" particles  reach  $\mathcal{I}^+$.

The Cauchy surface $\Sigma_{\infty}$ is the limit of the Cauchy surface $\Sigma_A$ of Fig. 1 as $A \rightarrow \iota^+$. Hence, the HW theorem can be viewed as a statement about the asymptotic form of the reduced density matrix defined with respect to the subset $C_1$ of $\Sigma_A$.  By construction, its conclusions are restricted to the outcomes of {\em single-time} measurements. The fact that multi-time measurements cannot be solely expressed in terms of a reduced density matrix is not affected by taking  one of the Cauchy surfaces to infinity.

For example,  two-time correlations may be   expressed in terms of the joint probability of detecting Hawking quanta by two apparatuses at two spacetime points in the Cauchy surfaces $\Sigma_A$ and $\Sigma_B$, as in the right-hand diagram of Fig. 1. The two detection events can have any separation  (timelike, spacelike, or null). They cannot be mapped to events on $\mathcal{I}^+$ without losing the key information of their causal relation.  As it has been shown in relation to the Unruh effect, multi-time correlations may well be non-thermal, even if all single-time properties are thermal
 \cite{AnSav11}.

We conclude that the probabilities of
multi-time  measurements  cannot be expressed solely in terms of the Gibbsian reduced density matrix (\ref{hawkrad}).
Hence, multi-time correlations in Hawking radiation are generically non-thermal, in the sense that they do not coincide with correlations obtained from a Gibbsian state.

\subsection{An open quantum systems perspective}
Next, we present  a more general argument why the HW theorem does not constrain multi-time correlations, based on well-known properties of {\em quantum open systems} \cite{Dav, BrePe}.

The HW theorem focuses on field properties at the future null infinity $\mathcal{I}^+$.
Of course, no physical measurements occur literally at   null infinity. The HW theorem is best viewed as a statement about the long time limit of an {\em open quantum system}.  Consider the Cauchy surface $\Sigma_A := (KA)\cup(A \iota_0)$ of Fig. 1, and the reduced density matrix $\hat{\rho}_A$ obtained by tracing out the degrees of freedom of the surface $KA$. The field degrees of freedom (dofs) on the surface  $A \iota_0$ are effectively an open system, with the dofs at the horizon playing the role of the environment.  Hence, the HW theorem is a  statement about  asymptotic thermalization   in an open quantum system.

 The key point is that in open quantum systems, the time-evolving reduced density matrix of a subsystem  does not contain all information about the subsystem. It contains only information accessible by single-time measurements. It does {\em not} contain sufficient information to correctly reproduce the probabilities of multi-time measurements, {\em unless the open system dynamics is Markovian} \cite{PaZu93, VA17}. Indeed, the idea that the single-time state contains all accessible information is an essential part of the definition of Markovian processes---see, for example, \cite{AFL, An03, RHP14}.

In non-Markovian processes, the environment keeps memory of properties of the system and releases this information to the system in a way that is not fully predictable by the open system dynamics.   {\em At the fundamental level},  open quantum systems that are defined by tracing out an environment have  non-Markovian dynamics. Markovian behavior emerges as a result of approximations. Hence, the HW theorem  only rules out asymptotic single-time correlations in the Hawking radiation. It does not rule out  temporal correlations, i.e., correlations defined in terms of   multi-time measurements.

\medskip

\section{Probability formulas for Unruh-DeWitt detectors}

\subsection{Multi-time measurements in QFT}
Multi-time measurements in QFT are not usually discussed in  particle physics,  where the emphasis is on the description of scattering experiments via $S$-matrix theory. In contrast, multi-time measurements  are ubiquitous in quantum optics. The joint detection probability of photons at different moments of time is essential for the definition of higher order coherences of the electromagnetic field, and for describing  phenomena like the Hanbury-Brown-Twiss effect,   photon bunching and anti-bunching \cite{QuOp}. These probabilities are usually constructed using Glauber's  photo-detection theory \cite{Glauber}. For a given quantum state $\hat{\rho}_0$ of the electromagnetic field, Glauber's theory expresses  the (unnormalized)  joint probability density $W_n(X_1, X_2, \ldots, X_n)$ for $n$ photodetection events at spacetime points $X_1, X_2, \ldots, X_n$ as
\begin{eqnarray}
W_n(X_1, X_2,\ldots, X_n) = Tr \left(\hat{E}^{i_n(+)}(X_n) \ldots \hat{E}^{i_2(+)}(X_2) \hat{E}^{i_1(+)}(X_1)  \hat{\rho}_0 \right.
\nonumber \\
\times \left. \hat{E}^{(-)}_{i_1}(X_1) \hat{E}^{(-)}_{i_1}(X_1)\hat{E}^{(-)}_{i_2}(X_2)\ldots \hat{E}^{(-)}_{i_n}(X_n) \right), \label{Glauber}
\end{eqnarray}
where $\hat{\bf E}^{(\pm)}(X)$ is the positive (negative) frequency part of Heisenberg-picture operators that represent the electric field strength.

Eq. (\ref{Glauber}) has a restricted domain of applicability: it presupposes that all detectors are at rest in a given frame, and it requires a non-local split of the field into positive- and negative-frequency components. This split could lead to non-causal behavior of the probabilities at large separations of the detectors.
Nonetheless, the simplicity and broad applicability of Glauber's theory render it into an important paradigm as a QFT measurement theory. Its crucial property is that the joint probability for $n$ detection events at different spacetime points is a linear functional of a specific QFT $2n$-point function.

In recent years, we  developed a new method \cite{AnSav12, AnSav13, AnSav17, AnSav19}  for describing  QFT measurements that shares the above property with Glauber's theory. We call this method the Quantum Temporal Probabilities (QTP) method, as its original motivation was to provide a general framework for  temporally extended quantum observables \cite{AnSav06, AnSav08, An08}. The key idea  is to distinguish between the time parameter of Schr\"odinger equation from the time variable associated to  particle detection \cite{Sav99, Sav10}. The latter time variable is then treated as a  macroscopic quasi-classical one  associated to the detector degrees of freedom. Hence, although  the detector is described  in microscopic scales by quantum theory, its macroscopic records are expressed  in terms of classical spacetime coordinates.

In  QTP, the interaction between  the field and the measurement apparatus is described by a local interaction Hamiltonian $\int d^3x \hat{O}(x) \hat{J}(x)$. In this expression, $\hat{O}(x)$ is a local composite operator for the field and $\hat{J}(x)$ is a current operator in the apparatus's Hilbert space. As a result, the probability density for $n$ measurement events, analogous to  Eq. (\ref{Glauber}), is a linear functional of the field $2n$-point function
\be
G(X_1, X_2, \ldots, X_n; X_1', X_2', \ldots, X_n') :=  Tr \left\{{\cal T}\left[ \hat{O}(X_n) \ldots \hat{O}(X_2)  \hat{O}(X_1)   \right] \hat{\rho}_0  \right.
\nonumber\\ \left.
\times  \bar{\cal T}\left[ \hat{O}(X_1') \hat{O}(X_2') \ldots  \hat{O}(X_n')  \right]  \right\},  \label{nmpt}
\ee
where ${\cal T}$ stands for time-ordering and $\bar{\cal T}$ for reverse-time-ordering.

The QTP probability assignment that features the correlation functions (\ref{nmpt}) is  constructed through a decoherent histories analysis of the measurement process \cite{Gri, Omn, GeHa}. In particular, it involves the identification of specific sets of histories associated to particle detection and the requirement that they satisfy approximate decoherence conditions. We note that  histories theory incorporates the quantum state reduction rule in its probability assignment.

In this paper, we employ a simpler description of measurements  by modeling the measuring apparatuses as  generalized UdW detectors. This model allows for an elementary and self-contained derivation of  a probability formula for multi-time probabilities that coincides with that of QTP in a particular regime. In particular, the probability assignment in the UdW models is obtained from the application of Born's rule for the detector's degrees of freedom, and it does not require a decoherent histories analysis.

\subsection{A single UdW detector}
A generalized UdW detector  is a pointlike quantum system that  follows a spacetime trajectory $X^{\mu}(\tau)$. Its internal degrees of freedom are described by a Hilbert space ${\cal H}$. Its self-Hamiltonian $\hat{h}$ generates time-translations with respect to the proper time parameter $\tau$. The detector  interacts with a quantum field $\hat{\phi}$, through a coupling  term $\hat{V}$ that takes the following form in the interaction picture

  \be
  \hat{V}(s) = \lambda \int d \tau  g_{\sigma} (s - \tau ) \int d^4 X \hat{O}(X) \delta^4(X - X(\tau)) \hat{m}(\tau), \label{Vint}
\ee
where  $\lambda$ is a coupling constant, $\hat{m}(\tau) = e^{i \hat{h}\tau} \hat{m} e^{- i \hat{h}\tau} $ for some operator $\hat{m}$ in the detector's Hilbert space, $\hat{O}(X)$ is a local composite operator of the field and  $g_{\sigma}(s)$ is a `switching' function, i.e.,  a function that determines when the  detector-field interaction is on. Our analysis holds for generic switching functions, however, we find convenient to employ   Gaussians

  \begin{eqnarray}
g_{\sigma}(s) = e^{- \frac{s^2}{2\sigma^2}}, \label{fsmear}
  \end{eqnarray}
with width $\sigma$. They satisfy the identity
\begin{eqnarray}
\sqrt{g_{\sigma}(\tau - s) g_{\sigma}(\tau - s')} =  g_{\sigma}(\tau - \frac{s+s'}{2}) g_{\sigma}(\frac{s-s'}{2}). \label{propertyf}
\end{eqnarray}

We assume that the self-Hamiltonian $\hat{h}$ of the detector has a unique ground state $|0\rangle$ at zero energy. We denote the other energy eigenstates as $|\epsilon \rangle$, for $\epsilon > 0$. The detector is initially (coordinate time $t = t_i$) prepared at $|0\rangle$ and the field at state $|\Psi\rangle$.  We consider a switching function centered at $\tau$, and we evaluate the probability $\mbox{Prob}(\epsilon, \tau)$ that the detector is found with energy $\epsilon$  at some coordinate  time $t = t_f$ after the coupling has been switched off. To this end, we employ the Dyson expansion for the associated evolution operator $\hat{U}(t_f, t_i)$
\begin{eqnarray}
\hat{U}(t_f, t_i) = \hat{I} +  \sum_{n=1}^{\infty} (-i)^n
\int_{t_i}^{t_f}  dt_1 \int_{t_i}^{t_1} dt_2 \ldots \int_{t_i}^{t_{n-1}} dt_n
\hat{V_c}(t_1) \hat{V_c}(t_2) \ldots \hat{V_c}(t_n), \label{timeordered}
\end{eqnarray}

 In Eq. (\ref{timeordered}),  $V_c(t): = \dot{\tau}(t) \hat{V}(\tau(t))$, where $\tau(t)$ is the inverse function of $X^0(\tau)$ that expresses $\tau$ as a function of $t$. It is necessary to express the interaction operator in terms of the  coordinate time $t$, because time ordering is defined with respect to the causal structure of spacetime, encapsulated in $t$, and not with respect to proper time parameters. This distinction is trivial for one detector, but it is essential for the system of multiple detectors that is examined later.
Since the switching-on functions vanish outside $[t_i, t_f]$, we can take $t_i \rightarrow -\infty$ and $t_f \rightarrow \infty$.

To leading order in the coupling constant $\lambda$,
\begin{eqnarray}
\mbox{Prob}(\epsilon, \tau) = \int d\tau g_{\sigma}(\tau - \tau') P(\epsilon, \tau'), \label{Prob0}
\end{eqnarray}
where
\begin{eqnarray}
P(\epsilon, \tau) = \alpha_\epsilon \int ds g_{\sigma}(\frac{s}{2}) e^{-i\epsilon s} G^{(2)}[X(\tau + \frac{s}{2}), X(\tau - \frac{s}{2})]. \label{prob1}
\end{eqnarray}
In Eq. (\ref{prob1}), $\alpha_\epsilon = \lambda^2 |\langle \epsilon|\hat{m}|0\rangle|^2$, and $G_2(X, X')$ is the field two-point function associated to the  composite operators $\hat{O}(X)$,
\begin{eqnarray}
G^{(2)}(X, X') = \langle \Psi|\hat{O}(X) \hat{O}(X')|\Psi\rangle.
\end{eqnarray}
The probability $\mbox{Prob}(\epsilon, \tau)$ is a density with respect to $\epsilon$, but not with respect to $\tau$. The proper time $\tau$ appears as a parameter in Eq. (\ref{Prob0}) and not as a random variable. In classical probability theory, we could define an unnormalized probability density $W(\epsilon, \tau)$ with respect to $\tau$ by dividing $\mbox{Prob}(\epsilon, \tau)$ with the effective duration of the interaction $T = \int ds g_{\sigma}(s) = \sqrt{2 \pi \sigma^2}$. Then, Eq. (\ref{Prob0}) would become
\begin{eqnarray}
W(\epsilon, \tau) = \int d\tau' f_{\sigma}(\tau - \tau') P(\epsilon, \tau'), \label{Prob0b}
\end{eqnarray}
where $f_{\sigma} = \frac{1}{\sqrt{2\pi \sigma^2}} e^{-\frac{x^2}{2\sigma^2}}$ is a probability distribution on $\mathbb{R}$. Hence,  the probability distribution $W(\epsilon, \tau) $ is the convolution of $P(\epsilon, \tau)$ with the smearing function $f_{\sigma}$.

The  definition (\ref{Prob0b}) of  density with respect to time is not rigorous for quantum probabilities, because it involves the combination of probabilities defined with respect to different experimental set-ups, i.e., different switching functions for the Hamiltonians. Nonetheless,   Eq. (\ref{Prob0b}) can  be derived from first principles in the context of the QTP method \cite{AnSav12, AnSav17, AnSav19}. The QTP method involves an explicit modeling of the interaction between the measuring apparatus using QFT. In particular, the composite operators $ \hat{O}(X)$ for $X = (t, {\pmb x})$ appear from a local  interaction term $\int d^3x \; \hat{O}({\pmb x} )\hat{J}({\pmb x})$, where $\hat{J}({\pmb x})$ is a current operator on the Hilbert space of the detector. In QTP, the smearing functions $g_{\sigma}$ are not interpreted in terms of a switching-on of the interaction, but they describe the {\em sampling} of a temporal observable associated to the value of the proper time $\tau$ at the instant of detection and incorporate the coarse-graining necessary for the definition of classicalized pointer variables in the apparatus.

For the purposes of this paper, it suffices to consider the probability density Eq. (\ref{Prob0}) derived from the elementary UdW detector model described above. Our conclusions about Hawking radiation do not depend  on the details of an elaborate quantum measurement description. However, the probability formulas can also be rigorously derived within perturbative QFT, in a way that avoids the main shortcoming of the UdW detector \cite{AnSav12, AnSav17, AnSav19}, namely, that it involves a coupling of particle and field degrees of freedom, that cannot be justified from first principles.

Two regimes are particularly important. If $\sigma$ is much smaller than all natural time scales of the system, then we may approximate $W(\epsilon, \tau) $ with $P(\epsilon,  \tau)$, i.e., treat $f_{\sigma}$ as a delta function. The opposite regime corresponds to very large $\sigma$, so that all information about the temporal localization of particle detection is lost. In this limit, $W(\epsilon, \tau)$ is time-independent and coincides with the time average of the probability $P(\epsilon, \tau)$.

 \subsection{A pair of detectors}
Next, we consider a pair of generalized Unruh-DeWitt detectors, moving along the spacetime paths $X_1^{\mu}(\tau_1)$ and $X_2^{\mu}(\tau_2)$. Each detector is described by a Hilbert space ${\cal H}_a$,  a self- Hamiltonian $\hat{h}_a$, and an interaction operator $\hat{m}_a$, where  $a = 1, 2$. The detectors couple to the field via  interaction terms of the form Eq. (\ref{Vint}) with composite operators $\hat{O}_a(X)$ and coupling constants $\lambda_a$. Hence, the interaction operator $\hat{V}_c(t)$ that enters the Dyson expansion is
\begin{eqnarray}
\hat{V}_c(t) =  \sum_{a=1}^{2} \lambda_a  \dot{\tau}_a(t)  \int ds g_{\sigma}^{(a)}[\tau_a(t- s)] \hat{O}_a[X_a(\tau_a(s))] \hat{m}_a(\tau(s)).
\end{eqnarray}

 For simplicity, we specialize to  detectors with identical internal characteristics, differing  only on their spacetime paths. Hence,  $\hat{h}_1 = \hat{h}_2 = \hat{h}$, $\hat{m}_1 = \hat{m}_2 = \hat{m}$, $\hat{O}_1(X) = \hat{O}_2(X) = \hat{O}(X)$, $\lambda_1 = \lambda_2 = \lambda$.

We assume that  both detectors are initially in their ground state.
We evaluate the joint  probability $\mbox{Prob}(\epsilon_1, \tau_1; \epsilon_2, \tau_2)$, that the first detector was switched on around  proper time $\tau_1$, the second detector was  switched on around  proper time $\tau_2$, and they were found with energies  $\epsilon_1$ and $\epsilon_2$ respectively after the interactions were switched off.
The leading contribution to the  probability amplitude comes from the second-order term in the Dyson expansion, hence,   $\mbox{Prob}(\epsilon_1, \tau_1; \epsilon_2, \tau_2)$ is of order $\lambda^4$. Again we take the limits $t_i \rightarrow -\infty$ and $t_f \rightarrow \infty$, since we assume that the coordinate time $t_i$ refers to a Cauchy surface prior the switching on of both detectors and $t_f$ refers to a Cauchy surface after both detectors have been switched off.

 It is convenient to work with the {\em coincidence function}
\begin{eqnarray}
C (\epsilon_1, \tau_1; \epsilon_2, \tau_2) := \frac{1}{2 \pi \sigma^2} \left[ \mbox{Prob}(\epsilon_1, \tau_1; \epsilon_2, \tau_2) - \mbox{Prob}(\epsilon_1, \tau_1) \mbox{Prob}(\epsilon_2, \tau_2) \right],
\end{eqnarray}
that quantifies the deviation from   uncorrelated detection. The coincidence function can be expressed as

\begin{eqnarray}
C (\epsilon_1, \tau_1; \epsilon_2, \tau_2)  = \int d\tau_1' d\tau_2' f_{\sigma}(\tau_1 - \tau_1') f_{\sigma}(\tau_2 - \tau_2') K (\epsilon_1, \tau_1'; \epsilon_2, \tau_2') , \label{coinc}
\end{eqnarray}
where
\begin{eqnarray}
K (\epsilon_1, \tau_1; \epsilon_2, \tau_2) = \alpha_{\epsilon_1} \alpha_{\epsilon_2} \int ds_1 \int ds_2 g_{\sigma}(\frac{s_1}{2}) g_{\sigma}(\frac{s_2}{2})  e^{- i\epsilon_1 s_1 - i\epsilon_2 s_2} \nonumber \\ \times G^{(4)}[X_1(\tau_1+\frac{s_1}{2}), X_2(\tau_2 + \frac{s_2}{2}), X_2(\tau_2 -  \frac{s_2}{2}), X_1(\tau_1 -  \frac{s_1}{2})]. \label{K12}
\end{eqnarray}
The field four-point function $G^{(4)}(X_1, X_2, X_1', X_2')$ is defined by
\begin{eqnarray}
G^{(4)}(X_1, X_2, X_1', X_2') :&=& \langle \Psi|\bar{\cal T}[\hat{O}(X_1)\hat{O}(X_2)] {\cal T}[\hat{O}(X_2') \hat{O}(X_1')]|\Psi\rangle
\nonumber \\
&-& G^{(2)}(X_1, X_1') G^{(2)}(X_2, X_2'),
\end{eqnarray}
where ${\cal T}$ stands for time-ordering and $\bar{\cal T}$ for reverse-time ordering.

 The four-point function $G^{(4)}$ involves mixed time-ordered and anti-time-ordered elements. It can similarly be shown that the probabilistic correlations for $n$ detectors are linear functionals of correlation functions $G^{(2n)}$ of $n$ time-ordered and $n$ anti-time-ordered entries. Such correlation functions do not appear in the usual formulation of QFT in terms of transitions between in and out states, i.e., the S-matrix formalism. They appear in
 the Closed-Time-Path (CTP) formulation of QFT, developed by Schwinger and Keldysh \cite{Schw, Kel, CTP}. This formulation allows one to derive probabilities for observables at any moment of time, and not only for asymptotic  properties.  The CTP correlation functions are obtained by
   functional differentiations of a  generating functional $Z[J_+, J_-]$,
 \begin{eqnarray}
 G^{(2n)}(X_1,\ldots, X_n, X'_1, \ldots, X'_n) =  \left(\frac{\delta^{2n}\ln Z[J_+, J_-]}{\delta J_-(X_1) \ldots \delta J_-(X_n) J_+(X'_1)\ldots \delta J_+(X'_n)} \right)_{J_{\pm}  = 0}.
 \end{eqnarray}
The CTP generating functional is defined as
\begin{eqnarray}
Z[J_+, J_-] := \langle \psi|\hat{U}^{\dagger}[J_-]\hat{U}[J_+]|\psi\rangle,
\end{eqnarray}
where $\hat{U}[J] = {\cal T} e^{-i \int d^4 X \hat{O}(X) J(X)}$, and ${\cal T}$ stands for time-ordered exponential.

In this paper, we will focus on QFTs with  Gaussian
CTP generating functional. This is the case for a free scalar field $\hat{\phi}(X)$ in a vacuum state, where the composite operators $\hat{O}(X)$ coincide with  $\hat{\phi}(X)$, or one of its derivatives.

For Gaussian generating functionals, the four-point correlation function $G^{(4)}$ is functionally determined by the two-point correlation functions $G^{(2)}$,
\begin{eqnarray}
G^{(4)}(X_1, X_2, X_1', X_2') = G^{(2)}(X_2, X_1') G^{(2)}(X_1, X_2') + G^{(2)}_F (X_2', X_1')  G^{(2)*}_F (X_2, X_1) \label{4ptgauss}
\end{eqnarray}
where
\begin{eqnarray}
G^{(2)}_F (X, X') = \theta(X^0 - X^{0\prime}) G^{(2)}(X, X') +  \theta(X^{0\prime} - X^0 ) G^{(2)}(X', X)
\end{eqnarray}
is the Feynman-two point function. For non-Gaussian generating functionals, an extra term should be added to Eq. (\ref{4ptgauss}), to account for the part of the four-point function that is functionally independent from the two point function.  For example, if $|\psi\rangle$ is chosen as the vacuum for a self-interacting scalar field, the additional term is standardly  evaluated through a perturbative expansion \cite{CTP}.

Eq. (\ref{4ptgauss}) implies that the   function (\ref{K12}) is a sum of two terms,
\begin{eqnarray}
K (\epsilon_1, \tau_1; \epsilon_2, \tau_2)   = K_0 (\epsilon_1, \tau_1; \epsilon_2, \tau_2) + K_F (\epsilon_1, \tau_1; \epsilon_2, \tau_2),
\end{eqnarray}
where
\begin{eqnarray}
 K_0 (\epsilon_1, \tau_1; \epsilon_2, \tau_2)  = \alpha_{\epsilon_1} \alpha_{\epsilon_2} \int ds_1 \int ds_2 g_{\sigma}(\frac{s_1}{2}) g_{\sigma}(\frac{s_2}{2}) e^{- i\epsilon_1 s_1 - i\epsilon_2 s_2} \nonumber \\
G(\tau_1 + \frac{s_1}{2}, \tau_2 - \frac{s_2}{2}) G^*(\tau_1- \frac{s_1}{2}, \tau_2 +  \frac{s_2}{2}) \label{K0} \\
K_F (\epsilon_1, \tau_1; \epsilon_2, \tau_2) = \alpha_{\epsilon_1} \alpha_{\epsilon_2} \int ds_1 \int ds_2 g_{\sigma}(\frac{s_1}{2}) g_{\sigma}(\frac{s_2}{2}) e^{- i\epsilon_1 s_1 - i\epsilon_2 s_2}  \nonumber \\
G_F(\tau_1 - \frac{s_1}{2}, \tau_2 - \frac{s_2}{2}) G_F^*(\tau_1+ \frac{s_1}{2}, \tau_2 +  \frac{s_2}{2}). \label{KF}
\end{eqnarray}

where we wrote
\begin{eqnarray}
G(\tau_1, \tau_2) : = G^{(2)}[X_1(\tau_1),  X_2(\tau_2)]\\
G_F(\tau_1, \tau_2) : = G^{(2)}_F[X_1(\tau_1),  X_2(\tau_2)].
\end{eqnarray}

Of particular interest is the   case of paths that generate {\em stationary correlations}, i.e., paths $X_i(\tau_i)$ such that $G(X_i(\tau_1),  X_j(\tau_2))$ is a function only of $\tau_1 - \tau_2$ for all $i, j = 1, 2$. We denote these functions by $\Delta_1$ if the entries in the two-point function correspond  to the same path, and by $\Delta_2$ if they correspond to different paths, i.e.,
\begin{eqnarray}
\Delta_1(s) = G(X_1(s),  X_1(0)), \hspace{2cm}
\Delta_2(s) = G(X_1(s),  X_2(0)).
\end{eqnarray}
We will call the functions $\Delta_1$ and $\Delta_2$ {\em correlators}.

 For such paths, the single-time probability density $W(\epsilon, \tau)$ of Eq. (\ref{prob1}) is time-independent
\begin{eqnarray}
W(\epsilon) = \alpha_{\epsilon} \int ds g_{\sigma}(\frac{s}{2}) e^{-i\epsilon s} \Delta_1(s). \label{PEc}
\end{eqnarray}
Furthermore, the   term $K_F$, Eq. (\ref{KF}), involves an integral
\begin{eqnarray}
\int dx e^{-\frac{x^2}{4\sigma^2} - i(\epsilon_1+ \epsilon_2)x} = \sqrt{4\pi \sigma^2} e^{-\sigma^2(\epsilon_1+\epsilon_2)^2}, \nonumber
\end{eqnarray}
 where $x = \frac{s_1+s_2}{2}$. Hence for $\epsilon_a \sigma >> 1$,  $K_F (\epsilon_1, \tau_1; \epsilon_2, \tau_2) \simeq 0$. The  term $K_0$, Eq. (\ref{K0}), involves an integral $\int dy e^{-\frac{y^2}{16\sigma^2} - i(\epsilon_1- \epsilon_2)y/2} = 4\sqrt{ \pi }\sigma^2 e^{- \sigma^2(\epsilon_1-\epsilon_2)^2}$, where $y = s_1-s_2$. For $\epsilon_a \sigma >>1$,    we can approximate $4\sqrt{ \pi \sigma^2} e^{- 2 \sigma^2(\epsilon_1-\epsilon_2)^2} =  4\pi \delta(\epsilon_1 - \epsilon_2)$. Then, $K (\epsilon_1, \tau_1; \epsilon_2, \tau_2) $ depends only on the difference $\tau_2  - \tau_1$,
\begin{eqnarray}
K (\epsilon_1, \tau_1; \epsilon_2, \tau_2)   = 4\pi   \alpha_{\epsilon_1} \alpha_{\epsilon_2}  \delta(\epsilon_1 -\epsilon_2) Q(\epsilon_1+\epsilon_2, \tau_1-\tau_2)   \label{K12f}
\end{eqnarray}
through the correlation function
\begin{eqnarray}
Q(\epsilon, \tau) = \int dx g_{\sigma}(x/\sqrt{2}) e^{- i \epsilon x}    \Delta_2(\tau  + x)\Delta_2^*(\tau - x).  \label{K12b}
\end{eqnarray}
Substituting Eq. (\ref{K12f}) into Eq. (\ref{coinc}) we find that the coincidence function $C(\epsilon_1, \tau_1;  \epsilon_2, \tau_2)$ depends only on $\Delta \tau : = \tau_1 - \tau_2$. Hence,
\begin{eqnarray}
C(\epsilon_1, \epsilon_2; \Delta \tau) = 4\pi   \alpha_{\epsilon_1} \alpha_{\epsilon_2}  \delta(\epsilon_1 - \epsilon_2)   \int d \tau' g_{\sigma}(\frac{1}{2}(\Delta \tau - \tau')) Q(\epsilon_1 + \epsilon_2, \tau'). \label{ce1e2}
\end{eqnarray}

\subsection{Energy coarse-graining}
So far,  we assumed maximal accuracy in the determination of energy. In a realistic macroscopic detector, energy must be coarse-grained. We can sample energy only with accuracy $\Delta \epsilon$, that must be much smaller than the recorded energies $\epsilon$, but must also satisfy $\Delta \epsilon \sigma >> 1$. To incorporate energy coarse-graining we convolute the probability density and coincidence functions defined above with a probability distribution $w(\epsilon)$ peaked around $0$ and with width equal to $\Delta \epsilon$, for example, a Gaussian $w(\epsilon) = (\sqrt{2\pi} \Delta \epsilon)^{-1} \exp\left(-\frac{\epsilon^2}{2(\Delta \epsilon)^2}\right)$.

As long as $\Delta \epsilon << \epsilon$ the changes to the probabilities are negligible, except for terms in the probability distributions that oscillate rapidly as $e^{i \epsilon L}$ for some length scale $L$, where $L >> (\Delta \epsilon)^{-1}$. Such terms are strongly suppressed. For the Gaussian smearing function $w(\epsilon)$, the suppression factor is $e^{-\frac{1}{2} L^2 (\Delta \epsilon)^2}$.

In what follows, we will refrain from writing energy coarse-graining explicitly in order to avoid cluttering our notation. We will take its contribution into account by dropping all suppressed oscillatory terms from the calculated probability distribution. Such terms appear   in the calculations of Sec. 4.4 and of the Appendix B.


\section{Temporal correlations in the Unruh effect  }

In this section, we revisit the correlations of accelerated detectors, first studied in Ref. \cite{AnSav11}. We
 consider a massless scalar field $\hat{\phi}(X)$ in Minkowski spacetime, and we  choose $\hat{O}(X) = \hat{\phi}(X)$. Then, the two point function $G^{(2)}(X, X')$ is the usual Wightman function.

 The Wightman function of a massless scalar field at a thermal state of temperature $T$  \cite{Wel00} is
\begin{eqnarray}
G^{(2)}_T(t, {\pmb x};t', {\pmb x}') = \hspace{9cm}
\nonumber \\
 - \frac{T \sinh(2\pi T|{\pmb x}-{\pmb x}'|)}{8 \pi |{\pmb x}-{\pmb x}'| \sinh(\pi T(t-t'-i\epsilon - |{\pmb x}-{\pmb x}'|)) \sinh(\pi T(t-t'-i\epsilon + |{\pmb x}-{\pmb x}'|)) },
\end{eqnarray}
where $\epsilon>0$ is taken to zero. For $T \rightarrow 0$, we obtain the vacuum Wightman function
\begin{eqnarray}
G^{(2)}_0(t, {\pmb x};t', {\pmb x}') = - \frac{1}{4\pi^2 [ (t-t'-i \epsilon)^2 - ({\pmb x}-{\pmb x}')^2]}.
\end{eqnarray}

We will compare the detection rate and correlations between   accelerated detectors in the  vacuum and static detectors in a thermal bath. First, consider a  detector at constant proper acceleration $a$.  The detector  moves along the trajectory    $X(\tau) = (a^{-1} \sinh(a \tau), a^{-1} (\cosh (a\tau) - 1) , 0, 0)$. The correlator   $\Delta^{(a)}_1(s) :=  G^{(2)}_0(X(\tau+s), X(\tau))$ is
\begin{eqnarray}
\Delta^{(a)}_1(s)  = -\frac{a^2}{16\pi^2\sinh^2\left(\frac{1}{2}a(s -i \epsilon)\right)}.
\end{eqnarray}

The correlator $\Delta^{(T)}_1(s) :=  G^{(2)}_T(X(\tau+s), X(\tau))$ for a static detector at  trajectory $X(\tau) = (\tau, 0, 0, 0)$  in a thermal bath of temperature $T$
is
\begin{eqnarray}
\Delta^{(T)}_1(s) = -\frac{T^2}{4\sinh^2\left(\pi T(s - i \epsilon)\right)}.
\end{eqnarray}
Obviously, $\Delta^{(T)}_1(s) = \Delta^{(a)}_1(s) $,  if  $T$ equals the Unruh temperature $T_U:= \frac{a}{2\pi}$. The detection rates given by Eq. (\ref{PEc}) are also identical.

Next, we consider  a pair of accelerated detectors moving along the spacetime trajectories
\begin{eqnarray}
X_1(\tau_1) &=& (a^{-1} \sinh(a \tau_1), a^{-1} (\cosh (a\tau_1) - 1), 0, 0)  \nonumber \\ \mbox{and} \;\;\;  X_2(\tau_2) &=& (a^{-1} \sinh(a \tau_2), a^{-1} (\cosh (a\tau_2) - 1), d, 0). \nonumber
\end{eqnarray}
 The trajectories are  separated by coordinate distance $d$, normal to the acceleration. The detector clocks are synchronized so that $X_1(0) = (0, 0, 0, 0,)$ and  $X_2(0) = (0, 0, d, 0,)$. The associated correlator
  $\Delta^{(a)}_2(s): = G^{(2)}_0(X_1(\tau+s), X_2(\tau))$ becomes
\begin{eqnarray}
\Delta^{(a)}_2(s) = -\frac{a^2}{16\pi^2\sinh\left(\frac{1}{2}a(s-i \epsilon - q) \right)\sinh\left(\frac{1}{2}a(s -i \epsilon + q)\right)}, \label{corr1}
\end{eqnarray}
where
\begin{eqnarray}
q = \frac{2}{a} \sinh^{-1}\left(\frac{ad}{2}\right)
\end{eqnarray}
is the proper time that it takes   a light signal to travel from one detector to  the other.

The paths for two static detectors at distance $q$ in a thermal bath  are $X_1(\tau_1) = (\tau_1, 0, 0, 0)$ and $X_2(\tau_2) = (\tau_2, 0, q, 0)$.  The correlator $\Delta^{(T)}_2(s): = G^{(2)}_0(X_1(\tau+s), X_2(\tau))$ is
\begin{eqnarray}
\Delta^{(T)}_2(s) = -  \frac{\sinh(2 \pi T q)}{2 \pi qT} \frac{T^2}{4 \sinh\left(\pi T(s -i \epsilon - q) \right)\sinh\left(\pi T(s-i \epsilon + q) \right)}. \label{corr2}
\end{eqnarray}
For $T = T_U$, the correlators  in Eqs. (\ref{corr1}) and (\ref{corr2})  are related by
\begin{eqnarray}
\Delta^{(a)}_2(s)  = \frac{\sinh aq}{aq} \Delta^{(T)}_2(s)
\end{eqnarray}
Hence, the coincidence function $C^{(a)} (\epsilon_1, \tau_1; \epsilon_2, \tau_2)  $ for a pair of accelerated detectors differs from the coincidence function $C^{(T)} (\epsilon_1, \tau_1;\epsilon_2, \tau_2)  $ for a pair of static detectors at the Unruh temperature by a distance-dependent factor
\begin{eqnarray}
C^{(a)} (\epsilon_1, \tau_1; \epsilon_2, \tau_2)  =  \left(\frac{aq}{\sinh aq}\right)^2C^{(T)} (\epsilon_1, \tau_1; \epsilon_2, \tau_2).
\end{eqnarray}

For $qa<< 1$, the two expressions almost coincide. Hence, the coincidences of accelerated detectors are thermal. However, for $qa >> 1$, they deviate significantly from thermal coincidences,  as they are suppressed by an exponential factor r $e^{-2aq}$.  In the Appendix B, we present a detailed calculation of $C^{(a)} (\epsilon_1, \tau_1; \epsilon_2, \tau_2)$ that  improves on the results of Ref.  \cite{AnSav11}.

Finally, we examine the case of two detectors separated along the direction of their acceleration. To this end, we consider trajectories
\begin{eqnarray}
X_1(\tau_1) &=& (a^{-1} \sinh(a \tau_1), a^{-1} \cosh (a\tau_1) - 1, 0, 0)   \nonumber \\  \mbox{and} \; \; \; X_2(\tau_2) &=& (a^{-1} \sinh(a \tau_2), a^{-1} \cosh (a\tau_2) - 1 + d, 0, 0). \nonumber
\end{eqnarray}
  The clocks are synchronized so that $X_1(0) = (0, 0, 0, 0,)$ and  $X_2(0) = (0, d, 0, 0,)$. In this case, the correlation function is non-stationary. The  correlation function
\begin{eqnarray}
G^{(2)}_0(X_1(\tau), X_2(\tau+s)) = \hspace{7cm} \nonumber \\
 -\frac{a^2}{16\pi^2\left(\sinh\left(\frac{1}{2}a(s -i \epsilon)\right) - dae^{-a\tau} \right)  (\sinh\left(\frac{1}{2}a(s -i \epsilon) + dae^{a\tau}) \right)}.
\end{eqnarray}
depends explicitly on $\tau$. This dependence cannot be removed even if we take $d$ to be a function of both $\tau_1$ and $\tau_2$. Again the coincidence function differs  from the one of   static detectors  in a thermal bath. Unlike thermal correlations, the correlations of accelerated detectors are not isotropic. Temperature is a spatial scalar  but acceleration is a spatial vector, and as such it defines a preferred direction.

In our opinion, the results above imply that the Rindler vacuum, obtained from Fulling-Rindler quantization \cite{Ful73}, cannot be taken literally as the quantum state of the field in the accelerated reference frame. In Fulling-Rindler quantization,
the restriction of the Minkowski vacuum in one Rindler wedge is a thermal state with respect to the Rindler time coordinate  associated to accelerated observers. If this state were interpreted as the quantum state of the field in the accelerated frame, one would expect that all   field observables with support on one Rindler wedge (including the correlations) would be distributed as if they were in a Gibbsian state.

However, the interpretation of the Rindler vacuum has no bearing on the--- fundamentally local---physics of the Unruh effect \cite{FulUn}. In particular, it does not challenge its thermodynamic character.
 The latter is established by the fact that microscopic systems undergoing constant acceleration in the Minkowski vacuum end up in a thermal state, regardless of their type of interaction with the quantum field \cite{MouAn, Mou2}. The results of Ref. \cite{AnSav11} and of this paper   suggest that this thermodynamic  characterization is not  valid for extended systems of size  $a^{-1}$ or larger. In contrast, a model for an extended quantum system was recently shown to also thermalize at the Unruh temperature \cite{LBHV}. Further analysis is required in order to settle this issue.

\section{Multi-time correlations in Hawking radiation}

In this section, we evaluate the coincidences for detectors in the Schwarzschild-Kruskal spacetime. We focus on the case of the Unruh vacuum, because it mimics the late-time  behavior of quantum fields in  black hole spacetimes that are formed in gravitational collapse \cite{Unruh76}.

\subsection{Preliminaries}

We consider a scalar field in the Schwarzschild manifold for a black hole of mass $M$. In the usual coordinate system $(t, r, \theta, \phi)$, and for $r > 2M$,
 the metric is
\begin{eqnarray}
ds^2 = - \left(1-\frac{2M}{r}\right) dt^2 + \frac{dr^2}{1-\frac{2M}{r}} + r^2 (d \theta^2 + \sin^2\theta d \phi^2).
\end{eqnarray}

The Regge-Wheeler  coordinate $r^* := r+2M \log\left(\frac{r}{2M}-1\right)$ becomes  $- \infty$ at the horizon and  $  \infty$ at spatial infinity. We also define the advanced radial coordinate $u : = t - r^*$ and the retarded radial coordinate $v = t + r^*$. The Kruskal coordinates, allowing for the maximal analytic extension of Schwarzschild spacetime, are $U = - \kappa^{-1}e^{-\kappa u}$ and $V = \kappa^{-1}e^{\kappa v} $, where $\kappa = (4 M)^{-1}$ is the surface gravity of the horizon.

Three quantum states have been proposed as vacua for QFTs on the maximally extended
Schwarzschild
manifold. These are
\begin{itemize}

\item the Boulware vacuum $|B\rangle$, defined in terms of normal modes that are positive frequency
with respect to the Killing vector $\frac{\partial}{\partial t}$ \cite{Boul75};

\item  the Unruh vacuum $|U\rangle$, defined in terms of incoming normal modes of the form $e^{-i \omega v}$ at ${\cal I}^-$ and  outgoing normal modes   of the form $e^{-i \omega U}$ at the past event horizon \cite{Unruh76};

\item the Hartle-Hawking-Israel vacuum $|H\rangle$, defined in terms of incoming modes that are positive frequency with respect to $V$, and
outgoing modes that are
positive frequency with respect to $U$ \cite{HaHa76, Isr76}.
\end{itemize}
The Wightman function associated to all three vacua is of the form \cite{ChFu77, Cand80}
\begin{eqnarray}
G(X, X') =  \sum_{\ell = 0}^{\infty} \int_{-\infty}^{\infty} \frac{d \omega}{4 \pi \omega} e^{- i \omega (t - t')} (2 \ell + 1)P_{\ell}({\pmb \Omega} \cdot {\pmb \Omega}') f_{\omega, \ell}(r, r'), \label{2pS}
\end{eqnarray}
where $X = (t, r, \theta, \phi)$, $X' = (t', r', \theta', \phi')$; ${\pmb \Omega}$ and ${\pmb \Omega}'$ are vectors on the unit sphere corresponding to $(\theta, \phi)$ and $(\theta', \phi')$, respectively; and $P_{\ell}$ are the Legendre polynomials. The functions $f_{\omega, \ell}(r, r')$  are constructed from the mode solutions to the  Klein-Gordon equation,
\begin{eqnarray}
\phi_{\omega, \ell, m} = \frac{1}{\sqrt{4 \pi \omega} r} R_{\omega, \ell}(r) Y_{\ell, m}(\theta, \phi) e^{- i \omega t},
\end{eqnarray}
where the radial functions
 $R_{\omega, \ell}(r)$  satisfy
\begin{eqnarray}
-\frac{d^2R_{\omega, \ell}}{dr^{*2}} + \left( 1 - \frac{2M}{r} \right) \left[ \frac{2M}{r^3} + \frac{\ell (\ell +1)}{r^2}\right] R_{\omega, \ell}  = \omega^2 R_{\omega, \ell}.  \label{radeq}
\end{eqnarray}
Eq. (\ref{radeq}) has the form of an one dimensional Schr\"odinger equation with a   potential
\begin{eqnarray}
V_{\ell}(r^*) = \left( 1 - \frac{2M}{r} \right) \left[ \frac{2M}{r^3} + \frac{\ell (\ell +1)}{r^2}\right]
\end{eqnarray}
that vanishes at $r^* \rightarrow \pm \infty$. The solutions  can be expressed  in terms of scattering theory in one dimension, and for $\omega^2 > 0$ they have double degeneracy. We   choose the  solutions $\overrightarrow{R}_{\omega, \ell}$ and $\overleftarrow{R}_{\omega, \ell}$, defined by their asymptotic behaviors
\begin{eqnarray}
\overrightarrow{R}_{\omega, \ell} (r) = \left\{ \begin{array}{cc} e^{i\omega r^*} + \overrightarrow{A}_{ \omega, \ell} e^{ - i \omega r^*}, & r_* \rightarrow - \infty\\ B_{\omega, \ell} e^{  i \omega r^*} & r^* \rightarrow  \infty\ \end{array} \right. \label{asy1}\\
\overleftarrow{R}_{\omega, \ell} (r) = \left\{ \begin{array}{cc} B_{\omega, \ell} e^{ - i \omega r_*} & r^* \rightarrow - \infty \\ e^{-i\omega r^*} + \overleftarrow{A}_{ \omega, \ell} e^{  i \omega r^*}, & r^* \rightarrow  \infty \end{array} \right.. \label{asy2}
\end{eqnarray}
 The functions $f_{\omega, \ell}(r, r')$ are different in each vacuum. For the Unruh vacuum \cite{Cand80},
 \begin{eqnarray}
 f_{\omega, \ell}(r, r') = \frac{1}{rr'} \left[ \frac{\overrightarrow{R}_{\omega, \ell} (r)\overrightarrow{R}_{\omega, \ell}^* (r')}{1 - e^{-\frac{2 \pi \omega}{\kappa}}} + \theta(\omega)\overleftarrow{R}_{\omega, \ell} (r)\overleftarrow{R}_{\omega, \ell}^* (r')\right], \label{fvvl}
 \end{eqnarray}
where we extend the definition of $R_{\omega, \ell}$ to negative $\omega$, by   $R_{-\omega, \ell} = R^*_{\omega, \ell}$. See, Ref.  \cite{Cand80} for the functions $f_{\omega, \ell}(r, r')$ in other vacua.

\subsection{The coincidence function}

In what follows we will consider paths $X(\tau) =  (\tau, r, \theta, \phi)$ that correspond to static detectors. The associated correlators are stationary when expressed in terms of the Killing time coordinate $t = \tau/ L(r)$ rather than the proper times $\tau$. For this reason, we will use the Killing time  as path parameter, and we will consider energies $E = L(r) \epsilon$ defined in terms of Killing-time translations.

For a pair of  detectors along the paths $X_1(t_1) = (t_1, r, \theta, \phi)$  and $X_2(t_2) = (t_2, r', \theta', \phi')$, the correlator is
\begin{eqnarray}
\Delta_2(s) = \int_{-\infty}^{\infty} \frac{d \omega}{4 \pi \omega} e^{-i \omega s} F_{\omega}(\Theta, r, r'), \label{deltachawk}
\end{eqnarray}
where $\cos \Theta = {\pmb \Omega} \cdot {\pmb \Omega}'$, and
\begin{eqnarray}
F_{\omega}(\Theta, r, r') = \sum_{\ell=0}^{\infty} (2 \ell + 1) P_{\ell}(\cos \Theta) f_{\omega, \ell}(r, r'). \label{Fvvr}
\end{eqnarray}

The correlator for a single path   $X(t) = (t, r, \theta, \phi)$ is
\begin{eqnarray}
\Delta_1(s) = \int_{-\infty}^{\infty} \frac{d \omega}{4 \pi \omega} e^{-i \omega s} F_{\omega}(0, r, r), \label{deltac1}
\end{eqnarray}

By Eq. (\ref{Prob0b}), the detection rate for a single static detector with $E  \sigma >> 1$ is constant
\begin{eqnarray}
W(E) = - \frac{\alpha_{E} }{2E} F_{-E}(0, r, r). \label{detecth1}
\end{eqnarray}
Using  Eq. (\ref{fvvl}), we recover the standard expression \cite{Cand80}
\begin{eqnarray}
W(E) = \frac{\alpha_{E} }{2E}  \frac{\sum_{\ell =0}^{\infty}(2\ell+1)  |\overrightarrow{R}_{E, \ell}(r)|^2}{e^{\frac{2\pi E}{\kappa}}-1}  . \label{detecth1b}
\end{eqnarray}

In particular, for a detector far from the black hole
\begin{eqnarray}
W(E) = \frac{\alpha_{E} T_E}{2E (e^{\frac{2\pi E}{\kappa}}-1)},
\end{eqnarray}
where $T_E = \sum_{\ell =0}^{\infty}(2\ell+1)  |B_{E, \ell}|^2$ is the escape probability from the potential.

The coincidence function $C (E_1, t_1; E_2, t_2)$ of Eqs.  (\ref{coinc}, \ref{K12})  is a function of  $\Delta t = t_1 - t_2$.
For $E_i \sigma >> 1$, we obtain the analogue of Eq. (\ref{ce1e2}),
\begin{eqnarray}
C (E_1, E_2; \Delta t) = 4 \pi \alpha_{E_1}\alpha_{E_2} \delta(E_1- E_2) Z(E_1, \Delta t),
\end{eqnarray}
where
\begin{eqnarray}
 Z(E,  \Delta t) &=& \frac{\sqrt{2}\sigma^2}{4\pi} \int \frac{d \omega_1 d \omega_2}{\omega_1 \omega_2} e^{- i(\omega_1 - \omega_2)\Delta t} e^{ - \sigma^2(2E +\omega_1 + \omega_2)^2 - 2 \sigma^2(\omega_1 - \omega_2)^2} \nonumber \\
 &\times&
F_{\omega_1}(\Theta, r, r') F_{\omega_2}^*(\Theta, r, r').
\end{eqnarray}

For $E \sigma >> 1$, both $\omega_1$ and $\omega_2$ take values very close to $-E$. Then, we can  approximate $\frac{\sigma}{\sqrt{\pi}} e^{ - \sigma^2(2E+ \omega_1 + \omega_2)^2} \simeq \delta (2E + \omega_1 + \omega_2)$ and also set $\omega_1 = \omega_2 = -E$ in the denominator.  Hence, the dominant contribution to  $Z(E,  \Delta t)$ is
\begin{eqnarray}
 Z(E,  \Delta t) = \frac{\sigma}{4 \sqrt{2\pi}E^2} \int d \xi e^{-2 \sigma^2 \xi^2- i \xi \Delta t} F_{-E + \frac{\xi}{2}}(\Theta, r, r')F^*_{-E - \frac{\xi}{2}}(\Theta, r, r'). \label{main4}
\end{eqnarray}

Eq. (\ref{main4}) is the main result of the section. It provides a closed expression for the coincidence function. It is valid for all three vacua described earlier. Here, we  focus on the Unruh vacuum.
An  exact evaluation  requires the knowledge of the solutions $R_{\omega, \ell}$, or equivalently of the Wightman function for the Schwarzschild spacetime. There are no known analytic expressions for these functions. However, an approximate expression for the Wightman function can, in principle, be derived  from a Gaussian approximation in the path integral \cite{BePa81, Page83}. Technically, the evaluation of the coincidence function is analogous to the calculation of the two point correlations of the stress tensor \cite{stresscor}, because they both involve the field four-point function. Unlike stress-energy correlations, the evaluation of the coincidence function does not require renormalization.

 In what follows, we will evaluate the coincidences in two  approximations. First, we will consider a simplified model that is often employed in demonstrations of Hawking radiation: we assume that the potential $V_{\ell}(r)$ vanishes, hence, there is no scattering. Second, we will examine the asymptotic behavior of the coincidences for $r^*, r^{*\prime} \rightarrow \pm \infty$ using a simplified description of one dimensional scattering. In the Appendix B, we present a formal calculation of the coincidence function through a stationary phase approximation.

\subsection{Effectively 2-d black hole}

We evaluate the coincidences Eq.  (\ref{main4}), by assuming that the potential vanishes. The mode functions are then $\overrightarrow{R}_{\omega, \ell} (r) = e^{i\omega r^*}$ and $ \overleftarrow{R}_{\omega, \ell} (r) = e^{-i\omega r^*}$. In this approximation, the angular degrees of freedom decouple and the field behaves as in a two-dimensional black hole.  In particular, the function $F_{\omega}(r, r')$ of Eq. (\ref{Fvvr}) becomes proportional to the delta function $\delta^2(\Theta) = \sum_{\ell = 0}^{\infty} (2\ell+1)P_{\ell}(\cos\Theta)$ on the unit two-sphere; in absence of scattering, the direction of propagation does not change. For the Unruh vacuum,
\begin{eqnarray}
F_{\omega}(\Theta, r, r') = \frac{\delta^2(\Theta)}{rr'} \left[ \frac{e^{i \omega (r^*-r^{\prime *})}}{1-e^{-\frac{2\pi \omega}{\kappa}}}+\theta(\omega) e^{-i \omega (r^*-r^{\prime *})}\right].
\end{eqnarray}
As in the associated two dimensional QFT, the corresponding Wightman function has an infrared divergence, and it needs to be regularized by introducing an infrared cut-off $\mu$.
The coincidence function depends on the regularization parameter $\mu$, but this dependence is of the order $\frac{E^2}{\mu^2} e^{- \sigma^2 E^2}$. Hence, it is strongly suppressed for $E \sigma >> 1$. In this regime, we can approximate $e^{-\frac{2\pi (E\pm \frac{1}{2}\xi)}{\kappa}}$ with $e^{-\frac{2\pi E}{\kappa}}$, to obtain
\begin{eqnarray}
Z(E,  \Delta t) = \frac{1}{8 E^2r^2 r^{\prime 2}}  \frac{g_{\sigma}(\frac{1}{2} \Delta u)}{(  e^{ \frac{2\pi E}{\kappa}}-1)^2} \delta^2(\Theta) \delta^2(0).  \label{zedt}
\end{eqnarray}
The formal divergence $\delta^2(0)$ arises because the calculation is essentially two-dimensional.

Particle detection events are correlated along the light-cone. They are non zero only if both events have the same coordinate $u$,  modulo the temporal accuracy $\sigma$ of the measurement. Only outgoing correlations are non-trivial, i.e., pairs of detection events where the detector closer to the black hole clicks first.
 Since  $C (E_1, E_2; \Delta t) > 0 $,  Hawking photons {\em bunch}, i.e., they tend  to be detected in pairs.


\subsection{Asymptotic behavior of coincidences}

We evaluate the coincidence function for the Unruh vacuum in the regime where the asymptotic expressions Eq. (\ref{asy1}, \ref{asy2}) for $r^* \rightarrow \pm \infty$ apply.   For $E \sigma >> 1$, the contribution from the $\overleftarrow{R}_{\omega, \ell} (r)$ solutions is negligible. In this section, we effect a rather drastic simplification of the scattering amplitudes that leads to simple results. The expressions we derive here are special cases of the formal expressions derived through a stationary phase approximation in the Appendix B.

We use the geometric optics approximation to Eq. (\ref{radeq}) that is valid in the limit of large frequencies. We assume that the wave is fully transmitted if the energy $\omega^2$ is larger than the maximum of the potential $V_{\ell}(r^*)$. The maximum is found at $r_0 = 2M/x_0$, where $x_0 = \frac{3}{8} (-\lambda +1+\sqrt{1+ \frac{14}{9}\lambda+\lambda^2})$, where $\lambda = \ell (\ell+1)$.  The value of the potential maximum is $V_{max}(\ell) = \kappa^2[2 \lambda - (\lambda - 1)x_0]x_0^2$. We denote by   $\ell_{c}(\omega/\kappa)$ the value of $\ell$ that solves the equation $\omega/\kappa = \sqrt{2 \lambda - (\lambda - 1)x_0}x_0$. $V_{max}(\ell)$ is well approximated by the linear function $\kappa^2(\frac{4}{3} \ell + \frac{3}{4})$, so $\ell_c (\omega/\kappa) =   \frac{3}{4} [(\omega/\kappa)^2 - \frac{3}{4}]$. This implies that for   $\omega/\kappa < \frac{\sqrt{3}}{2}$, no particle crosses the barrier. Obviously, the geometric optics approximation is unreliable at  low energies, including the  most interesting regime of $\omega \sim \kappa$. However, it suffices for demonstrating the behavior of the coincidence function.

In the geometric-optics approximation,
\begin{eqnarray}
\begin{array}{ccc}  |B_{\omega, \ell}| = 1, & \mbox{and} \; \;  |\overrightarrow{A}_{ \omega, \ell}| = 0 &  \mbox{for} \; \;  \ell \leq \ell_{c}(\omega/\kappa) \\ |B_{\omega, \ell}| = 0, & \mbox{and} \; \; |\overrightarrow{A}_{ \omega, \ell}| = 1  &  \mbox{for} \; \;  \ell > \ell_{c}(\omega/\kappa). \end{array}
 \end{eqnarray}
 The maximum of the potential is very close to $r = 3M$, with less than $2 \%$ deviation for all $\ell > 1$. Hence,  we can take $r = 3M$ as the point of reflection for $\ell > \ell_c$. This corresponds to $r^* = \bar{r}^* = (3 - 2 \ln 2)M \simeq  1.61M$. Requiring that $\overrightarrow{R}_{\omega, \ell} (r = 3M) = 0$, we obtain $\overrightarrow{A}_{ \omega, \ell} = -e^{2 i\omega \bar{r}^* }  $ for $\ell > \ell_{c}$.

 We  evaluate the  Eq. (\ref{main4}) at the limit  $E \sigma >>1$, where we can approximate $e^{\frac{2\pi (E\pm \frac{1}{2}\xi)}{\kappa}}$ with $e^{\frac{2\pi E}{\kappa}}$. We obtain the following.

\medskip

\noindent {\bf Case I:} both detectors are far from the horizon. We find that

\begin{eqnarray}
Z(E,  \Delta t) = \frac{1}{8 E^2r^2 r^{\prime 2}}  \frac{g_{\sigma}(\frac{1}{2} \Delta u)}{(  e^{ \frac{2\pi E}{\kappa}}-1)^2} T_E(\Theta)^2, \label{zedt2}
\end{eqnarray}
 where
 \begin{eqnarray}
 T_E(\Theta) = \sum_{\ell = 0}^{\ell_c(E/\kappa)} (2 \ell +1)  ^2 P_{\ell}(\cos \Theta).
 \end{eqnarray}
 The result has the same form with Eq. (\ref{zedt}), modulo the angular dependence encoded in the function $T_E(\Theta) $. However, there is a significant difference. In four dimensions, the correlations are not along a null geodesic that connects the two detection events, since the condition $\Delta u = 0 $ ignores the angular coordinates. In fact, curves that satisfy $\Delta u = 0 $ are spacelike---unless $\Theta = 0$, in which case they are null. Hence, correlations between Hawking quanta are non-causally connected.

  \medskip

\noindent  {\bf Case II: }both detectors are near the horizon.    For this case, we define $\Delta r^*_{ref} := r^* - (2\bar{r}^* - r^{\prime *})$, the difference in distance traveled between one outcoming Hawking quantum detected at $r^*$ and one  incoming Hawking quantum that was reflected at $\bar{r}^*$ and then detected at $r^{\prime *}$. We also define $\Delta u_{ref} := \Delta t - \Delta r^*_{ref}$ and $\Delta v_{ref} := \Delta t + \Delta r^*_{ref}$, where the index makes reference to a reflected  quantum.

Then, Eq. (\ref{main4}) gives
\begin{eqnarray}
Z(E,  \Delta t) = \frac{1}{8 E^2r^2 r^{\prime 2}(  e^{ \frac{2\pi E}{\kappa}}-1)^2} \left\{ g_{\sigma}(\frac{1}{2} \Delta u) \left[\delta^2(\Theta)\right]^2
\nonumber  \right. \\
\left.
+ 2 \cos (2 E \Delta r_*) g_{\sigma}(\frac{1}{2} \Delta t) \delta^2(\Theta) S_E(\Theta)
\nonumber  \right. \\
\left.
+ S_E(\Theta)^2 \left[ g_{\sigma}(\frac{1}{2} \Delta v) + g_{\sigma}(\frac{1}{2} \Delta u^{ref}) + g_{\sigma}(\frac{1}{2} \Delta v^{ref})\right]
\right\},
 \label{zedt3}
\end{eqnarray}
where
\begin{eqnarray}
S_E(\Theta) = \sum_{\ell=\ell_c(E/\kappa)+1}^{\infty} (2 \ell + 1)   P_{\ell}(\cos \Theta) = \delta^2(\Theta) - T_E(\Theta).
\end{eqnarray}
In the derivation of Eq. (\ref{zedt3}), we ignore rapidly oscillatory terms that are suppressed after coarse-graining of energy---see, Sec. 3.4.

The first  term in Eq. (\ref{zedt})    refers to a pair of outcoming Hawking quanta.
The second term is non-zero  only for $\Delta t = 0$, irrespective of the distance between the two detectors. This term is multiplied by an oscillating factor that persists after  energy coarse-graining at a scale $\Delta E$, unless $\Delta r^* >> (\Delta E)^{-1}$.

The third term corresponds to a pair of Hawking quanta that are both incoming after reflection. The last two terms corresponds to pairs of Hawking quanta, such that one is detected while outgoing, and the other is detected when incoming    after reflection.

The approximations employed here fail for detectors arbitrarily close to the horizon. The problem is not so much with the geometric optics approximation, as with the condition
 $E \sigma >> 1$; the reason is that $E \rightarrow 0 $ on the horizon. Given any large negative $r^*$,  $V_{\ell}(r^*) $ is significant at $r^*$, hence, the asymptotic form of the solution is not applicable at $r^*$. The inclusion of all values of $\ell$ leads to  the divergent term $[\delta^2(\Theta)]^2$.

\medskip

\noindent {\bf Case III:} one detector near and one far from the horizon. In this case, we recover Eq. (\ref{zedt2}). In the geometric optics approximation, there is no interference between different partial waves when crossing the barrier  and no correlation between escaped and reflected quanta. In the general case, such terms exist---see, the Appendix B.

 \medskip

 We conclude that there are significant multi-time correlations in Hawking radiation. They are strongly dependent on the potential $V_{\ell}(r)$ and on the angular separation of the two events. In contrast,  for single-time measurements, the effect of the potential is contained in a single function that corresponds to the escape probability. Even with  the rough approximation effected here, we can see that correlation measurements reveal significant information about the history of the Hawking quanta, especially for detectors   close to the horizon. The more elaborate analysis of Appendix B reveals further degrees of complexity.

\section{Physical interpretation}
We showed that there are non-trivial multi-time correlations in Hawking radiation, and that their form is not constrained by the HW theorem. We evaluated these correlations using a simple method based on generalized   UdW detectors. In this section, we discuss the physical implications of our results.

\medskip

\noindent {\em Black hole thermodynamics.} Our results reaffirm the thermodynamic character of black holes. We showed that the   thermal behavior of Hawking radiation is manifested {\em only} in single-time measurements, which access the information of the field two point function. Multi-time measurements access information from higher-order field correlation functions, and they  do not manifest thermal behavior.
Hence, the thermodynamic description of Hawking radiation requires {\em coarse-graining} at the level of the two-point function, i.e., the same coarse-graining that defines the thermodynamic level of description for matter thermodynamics\footnote[3]{To see this, note that for non-relativistic fields the restriction to single-time measurements, and hence, to two-point functions, is equivalent to a description in terms of a single-particle reduced density matrix \cite{AnSav17}, through which the quantum Boltzmann equation is defined. For relativistic fields, the restriction leads to the Kadanoff-Baym equation \cite{KaBa}, or to the relativistic Boltzmann equation \cite{CTP, SLG}.}. We believe that this point is crucial for understanding the origins of the Generalized Second law in black hole thermodynamics.

\medskip

\noindent {\em Recovering pre-collapse information.} Since there is no physical  mechanism to block black hole radiation, a black hole will continuously lose energy, until it evaporates completely---see, Fig. 2. At the end of this process, all mass of the black hole will have been emitted as  radiation. Since the horizon will have disappeared, the reduced density matrix of Hawking quanta will be the full density matrix for matter, and it will be thermal according to the HW theorem. Hence, if we consider the full process of collapsing matter, black hole formation and evaporation, we end up with a mixed state even if the initial state of the total system of matter and gravity is pure. This implies a non-unitary evolution law that takes pure states to mixed ones \cite{Hawk76}, hence, loss of information. This conclusion  is often regarded  as unacceptable, whence one refers to the above argument as the "paradox"  of black hole information loss.

Researchers looking for a restoration of unitarity   often use the heuristic image of information storage, and they inquire where the missing information could be stored. The HW theorem is usually taken to imply that information cannot be stored in Hawking radiation correlations.  The usual argument is the following. Even if we include the effects of backreaction from the quantum field, one expects that the black hole geometry changes in a quasi-stationary way, so that the geometry can   be well approximated by a Schwarzschild solution with a slowly changing mass.
Hence, the semiclassical approximation will be good until the black hole shrinks near the Planck mass.  Quantum gravity effects at this stage  cannot affect the radiation that has already been emitted.  It follows that
the final state cannot be very different from thermal, i.e., highly mixed and with a limited capacity to carry information.


 \begin{figure}
    \centering

 \includegraphics[width=0.45\textwidth]{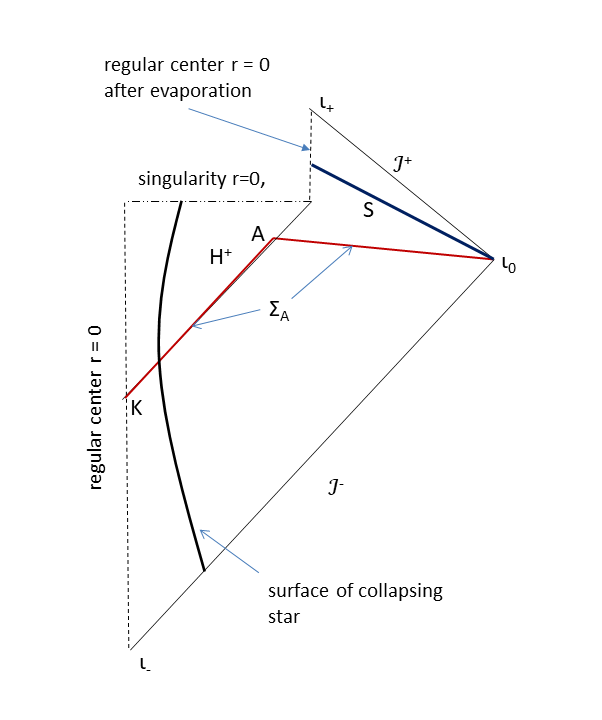}

    \caption{ Penrose diagram for black hole collapse and total evaporation. }
    \label{fig:foobar}
\end{figure}

Our work shows that, on the contrary,  correlations associated to multi-time measurements of the quantum field are non-trivial: they have a sufficiently complex form, and they keep memory of processes involving the scattering of Hawking quanta---this is shown in Sec. 5.4 and in the Appendix B.
Hence, multi-time correlations can carry significant amount  of information. This information does not amount to a small modification of  thermal behavior---as, for example,  when considering the effect of a non-vacuum field initial state \cite{LP1, LP2}. As in the simpler analysis of the Unruh effect in Sec. 4, there are multi-time measurements with the probabilities strongly deviating   from those of a Gibbsian state. Even though we studied the special case of Schwarzschild spacetime, there is nothing special in the    behavior of multi-time correlations found here. We expect that an analysis of the Kerr-Newman family of solutions will show that the correlations have an even more complex form.

For the reasons above, we find it highly plausible that  backreaction stores some pre-collapse information in multi-time correlations.
This storage process requires no new physics and no input from quantum gravity, because it  has a clear analogue in non-equilibrium statistical mechanics. In that context, information is lost to the thermodynamic  level of description because it is transferred to thermodynamically  inaccessible degrees of freedom, including non-local correlations. Hence, thermodynamical entropy increases. This analogy between black-hole information loss and inaccessible information in non-equilibrium statistical mechanics has been pointed out by Page \cite{Page93} and   by Calzetta and Hu \cite{CaHu95}.

The argument above becomes clear by considering backreaction expressed in terms of  perturbations around the classical geometry.   According to the Schwinger-Dyson equations, the
coupling of the scalar field to gravitational perturbations, either classical or quantum, is an interaction channel through which information is transferred to higher-order correlation functions of the scalar field. This information cannot be retrieved from quantities that depend solely on the two-point function. One such quantity is
the  expectation value of the stress-tensor, that defines the level of description  relevant to  back-reaction. The missing information
  is dispersed over all spacetime,
 and this is why it can only be accessed by multi-time measurements, i.e., measurements that are not localized in a single spacetime region.

  The transfer of information to inaccessible degrees of freedom is a continuous and persistent  process that does not require significant transfer of energy. This means that   backreaction could dramatically change the information balance in the higher order correlation functions, without significantly affecting the expectation value of  the stress-energy tensor. Hence, there is no conflict: pre-collapse information can be contained in higher order correlation functions while the  semi-classical approximation remains adequate until the later stages of black hole evaporation.

 In future work, we will test the above conjecture in simple backreaction models. We do not expect unitarity to be restored,  even if the conjecture proves correct. We think that the issues of unitarity and of information survival are conceptually distinct, in particular, the latter does not imply the former.

 Our investigation into the retrieval of pre-collapse information is primarily  motivated by the possibility that multi-time correlations could define {\em quantum informational hair} for a black hole.
 By this we mean the following. In   General Relativity, the no-hair theorem asserts that the black hole keeps no information from the initial state except for mass, angular momentum and charge. At the level of QFT in curved spacetime, the long-time behavior of the quantum field is accurately described by the Unruh vacuum that also carries no memory of the initial state except for mass, angular momentum and charge. In this sense, the universality of the Unruh vacuum is a quantum manifestation of the classical no-hair theorem\footnote[3]{The universality of the Unruh vacuum is well  accepted, even though there is no general proof---as far as we know.
  We believe that the Unruh vacuum exhibits this universality also when multi-time correlations are taken into account. We analysed the simple models for gravitational collapse  of Refs. \cite{Unruh76, BiDa},  and found that multi-time correlations after the formation of the
    horizon (not necessarily late times) are strongly dominated by  the Unruh vacuum terms. Hence, no significant pre-collapse information is stored in multi-time correlations {\em prior} to the incorporation of backreaction.}. The question  is whether this  loss of information of the initial state persists after the inclusion of backreaction.

    The no-hair property will be true  after   backreaction, if we can prove that the probabilities of multi-time measurements at late times are obtained by a unique generating functional that does not depend on pre-collapse properties of the system. In contrast, if the results of multi-time measurements after the collapse allows us to {\em retrodict}   properties of the system before collapse in a mathematically rigorous way, then  quantum informational   hair exist.

     \medskip

\noindent {\em Unitarity and information.} Our perspective about  non-unitarity in the black hole evaporation process coincides  with that of Unruh and Wald \cite{UnWa}. We view    the non-unitarity  of quantum gravity as a {\em prediction} of  the semi-classical QFT analysis, rather than a paradox or a breakdown of quantum theory. Non-unitarity originates from the fact that an instant of `time'  after evaporation, i.e., the spacelike surface $S$ of Fig. 2, is not  a Cauchy surface. In fact, the Kodama-Geroch-Wald theorem implies that even  the surface $\Sigma_A$ of Fig. 2   fails to be  Cauchy \cite{KGZ}. In other words, the spacetime develops `pathologies'  even before evaporation.

We think that it is more accurate to talk about the breakdown of the notion of a single-time quantum state, rather than violation of unitarity. After all,   single-time states---or equivalently, evolving single-time observables, as in the Heisenberg picture---are inseparably linked to Cauchy surfaces both in classical and in quantum field theory.
This suggests that generalizations of quantum theory that  are based on the notion of history \cite{GeHa, Ish94, Sor94}--- treating single-time quantum states as derived concepts---are more appropriate for the physics of black hole evaporation \cite{hartle98}, and arguably, for quantum gravity \cite{hartlelo, Sor94b, Sav03, Sav10}.

  The multi-time measurements described here fit naturally with a histories description. The associated probabilities can be defined in terms of history variables and   the decoherence functional \cite{AnSav12, AnSav17, AnSav19}, i.e., the mathematical object that generalizes the notion of the quantum state and incorporates probabilities in histories theory. Hence, they remain meaningful notions even  in non-globally hyperbolic spacetimes, such as the evaporating black hole spacetime of Fig. 2.

Multi-time probabilities   incorporate novel notions of quantum information that are not accessible in the description of a system in terms of single-time quantum states. For example, entanglement refers specifically to single-time correlations, it cannot be employed for correlations between spacetime regions that are not spacelike separated. In fact, the heuristic image of information being stored `somewhere' is misleading for  such correlations.
We believe that a covariant generalization of existing quantum information concepts is crucial for understanding information in relativistic systems.

\section*{Acknowledgements}
CA acknowledges support by Grant No. E611 from the Research Committee of the University of Patras via the “K. Karatheodoris” program.


\begin{appendix}

\section{Evaluation of coincidences in accelerated detectors}
We calculate the coincidence function (\ref{ce1e2}) for a pair of accelerated detectors using the correlation function (\ref{corr1}). We consider three regimes.

\medskip

\noindent {\bf Case I: $d = 0$}.  The Planckian detection spectrum arises for  $a \sigma >> 1$ \cite{AnSav11}. To leading order in $(\sigma a)^{-1}$, $Q(\epsilon, \tau)$ is independent of $\sigma$,
\begin{eqnarray}
Q(\epsilon, \tau) &=& \frac{a^4}{256 \pi^4} \int_{-\infty - i \eta}^{\infty - i \eta} \frac{dx e^{-i\epsilon x}}{\sinh^2[\frac{1}{2}a(x-\tau)] \sinh^2[\frac{1}{2}a(x-\tau)] }
\nonumber \\
&=&\frac{a^2 }{64 \pi^2 (e^{2\pi \epsilon/a} - 1)} \frac{ a \coth a\tau \sin \epsilon\tau- \epsilon \cos \epsilon \tau }{\sinh^2a\tau}
\end{eqnarray}

The physically relevant regime corresponds to energies such that $\epsilon \sigma >> 1$.  In Eq. (\ref{ce1e2}), $Q(\epsilon, \tau)$ is peaked around $\epsilon = 0$ and it oscillates with $\epsilon \tau$. For  $\epsilon \sigma >> 1$, we can
remove the dependence of $g_{\sigma}$ on $\tau'$ in Eq. (\ref{ce1e2}), to obtain the following expression for the coincidence function
\begin{eqnarray}
C(\epsilon_1, \epsilon_2; \Delta \tau) =  4\pi   \alpha_{\epsilon_1} \alpha_{\epsilon_2}  \delta(\epsilon_1 - \epsilon_2)  g_{\sigma}(\frac{1}{2}\Delta \tau ) \bar{Q}(\epsilon_1+ \epsilon_2),
\end{eqnarray}
where
\begin{eqnarray}
\bar{Q}(\epsilon ) := \int_{-\infty}^{\infty} d \tau  Q(\epsilon , \tau) = \frac{\epsilon^2 \coth \frac{\pi \epsilon}{2a}}{128 \pi (e^{\frac{2\pi \epsilon}{a}} - 1)}.
 \label{QE1}
\end{eqnarray}
In evaluating Eq. (\ref{QE1}), we employed the integral $\int_0^{\infty} dx ( b \coth bx \sin x- \cos x)/\sinh^2(bx) = \frac{\pi}{4b^2} \coth\frac{\pi}{2b}$.

\medskip

\noindent
{\bf Case II: $\sigma << q$.}
For finite $d$, Eq. (\ref{K12b}) becomes
\begin{eqnarray}
Q(\epsilon, \tau) =  \frac{a^4}{256 \pi^4} \int_{-\infty-i \eta}^{\infty - i \eta} \frac{dx g_{\sigma}(x/\sqrt{2}) e^{- i \epsilon x}}{\sinh[\frac{1}{2}a(x-u)]  \sinh[\frac{1}{2}a(x+u)]  \sinh[\frac{1}{2}a(x - v)] \sinh[\frac{1}{2}a(x+v)]  }. \label{qet}
\end{eqnarray}
where $u = \tau  - q$ and $v = \tau +q$.

We examine the case that the detector resolution is much smaller than the time it takes a signal from one detector to the other, $\sigma << q$. In this regime, the peaks at $|x| = v$ and $|x| = u $ do not overlap. Then, for $|\tau| >> \sigma$,  we approximate
\begin{eqnarray}
Q(\epsilon, \tau) =  \frac{a^4}{256 \pi^4\sinh(a\tau) \sinh(aq) }\nonumber \\
\times \left[
 g_{\sigma}(u/\sqrt{2}) \int_{-\infty-i \eta}^{\infty - i \eta} \frac{dx e^{- i \epsilon x} }{\sinh[\frac{1}{2}a(x - u)] \sinh[\frac{1}{2}a(x+u)}
\right. \nonumber \\
\left. -  g_{\sigma}(v/\sqrt{2}) \int_{-\infty-i \eta}^{\infty - i \eta} \frac{dx e^{- i \epsilon x} }{\sinh[\frac{1}{2}a(x - v)] \sinh[\frac{1}{2}a(x+v)]  }  \right] .
 \label{qet2}
\end{eqnarray}
We carry out the integration, to obtain
\begin{eqnarray}
Q(\epsilon, \tau) = \mbox{sgn}(q) \frac{a^3 e^{-a (|\tau| + |q|)}}{8\pi^3(e^{\frac{2\pi \epsilon}{a}}+1)  } \left( g_{\sigma}(u/\sqrt{2})  \frac{\sin \epsilon u}{\sinh a u} - g_{\sigma}(v/\sqrt{2})\frac{\sin \epsilon v}{\sinh a v}\right), \label{qet3}
\end{eqnarray}
In Eq. (\ref{qet3}), we approximated $\sinh(a \tau) \sinh (a q) \sim \mbox{sgn}(q) \frac{1}{4}e^{-a (|\tau| + |q|)}$, since $a \tau >> a \sigma >> 1$ and $a |q| >> a \sigma >> 1$.

By Eq. (\ref{ce1e2}), the coincidence function involves two separate contributions. The first comes from oscillations with frequency $\epsilon_1 + \epsilon_2$ around $\tau = q$, and the second from oscillations with same  frequency around $\tau = -q$. We evaluate each term separately. We obtain an analytic approximation valid for $\epsilon \sigma >> 1$, by substituting  $e^{-a |\tau|}$ with  $e^{-a |q|}$, and removing the $\tau'$ dependence from the $g_{\sigma}$ terms,
\begin{eqnarray}
C(\epsilon_1, \epsilon_2; \tau) = \mbox{sgn}(q)|\alpha_{\epsilon_1}|^2\frac{a^2 \tanh\left(\frac{\pi \epsilon_1}{a}\right)}{2\pi (e^{\frac{4\pi \epsilon_1}{a}} + 1) }e^{-2 a|q|} \delta(\epsilon_1 - \epsilon_2)  \nonumber \\ \times \left( g_{\sigma}(\frac{\Delta \tau - q}{\sqrt{2}})  - g_{\sigma}(\frac{\Delta \tau + q)}{\sqrt{2}}) \right),
\end{eqnarray}
where we employed the integral $\int_{-\infty}^{\infty}dx \sin(bx)/\sinh(x)= \frac{\pi}{2} \tanh\frac{b\pi}{2}$.

\medskip
\noindent
{\bf Case III: $\sigma \rightarrow \infty$.} We set $g_{\sigma} = 1$ in Eq. (\ref{qet}). Then,
\begin{eqnarray}
Q(\epsilon, \tau) = \frac{a^3 }{32\pi^3(e^{\frac{2\pi \epsilon}{a}}+1) \sinh aq \sinh a \tau  } \left(  \frac{\sin \epsilon (\tau - q)}{\sinh a (\tau -q)} - \frac{\sin \epsilon (\tau + q)}{\sinh a (\tau + q)}\right), \label{qet5}
\end{eqnarray}
The coincidence function  is $\tau$ independent
\begin{eqnarray}
C(\epsilon_1, \epsilon_2; \tau) = |\alpha_{\epsilon_1}|^2 \frac{a^2 \tanh\left(\frac{\pi \epsilon_1}{a}\right)}{4\pi^2 (e^{\frac{4\pi \epsilon_1}{a}} + 1) }\delta(\epsilon_1 - \epsilon_2) {\cal M}( qa, \epsilon/a),
\end{eqnarray}
where
\begin{eqnarray}
{\cal M}(x, b) : = \frac{1}{\sinh x} \int_0^{\infty} \frac{dy}{\sinh y} \left( \frac{\sin b(y-x)}{ \sinh (y-x)} - \frac{\sin b(y+x)}{ \sinh (y+x)} \right). \label{funcM}
\end{eqnarray}
In Fig. 3, ${\cal M}$ is plotted as a function of $x$ for different values of $b$. Since the measurement cannot resolve the delayed propagation of signals between the detectors,  ${\cal M}$ is peaked around $q = 0$ for all $b$ with a width of order $a^{-1}$.

\begin{figure}[]
\includegraphics[height=9cm]{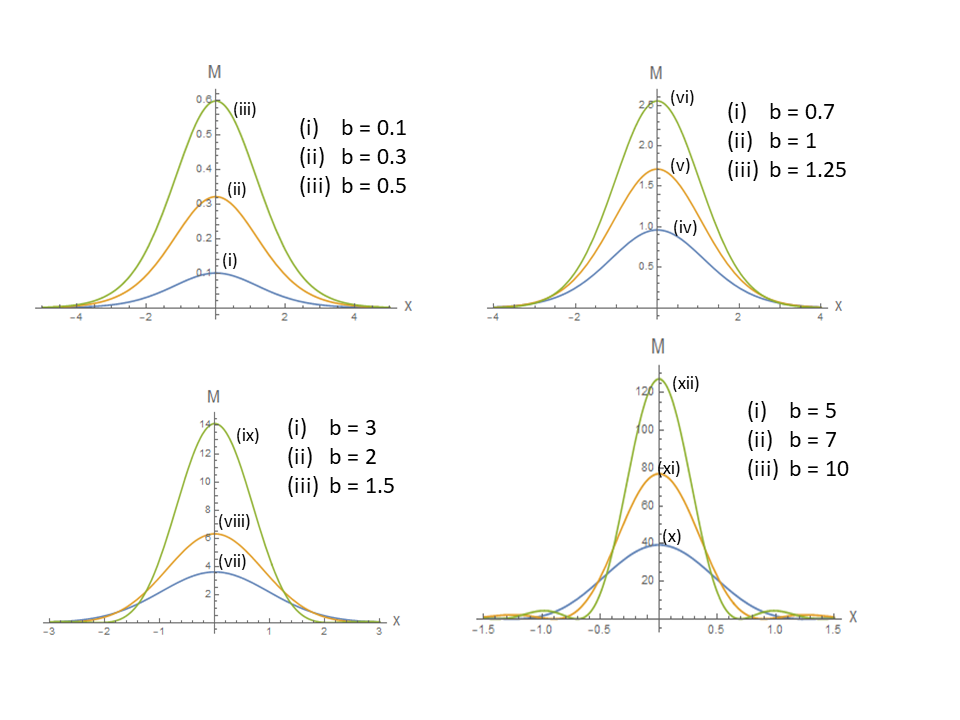} \caption{ \small The function ${\cal M}$, defined by Eq. (\ref{funcM}) is plotted as a function of $x$ for different values of $b$.}
\end{figure}

\section{Stationary phase evaluation of correlations in Hawking radiation}
In this section, we
 evaluate the coincidence function for the Unruh vacuum in the regime where the asymptotic expressions (\ref{asy1}, \ref{asy2}) for $r_* \rightarrow \pm \infty$ apply.  Since the integral (\ref{main4}) is strongly dominated by values of $\omega$ around $-E$, we employ the usual stationary phase approximation for wave-packet propagation in one-dimensional scattering. That is, we expand   $\ln \overrightarrow{A}_{ \omega, \ell}$ and $B_{\omega, \ell}$ as a series around $\omega = -E$,  keeping only the zero-th order term for the real part and up to the first-order term for the imaginary part.

    Hence, we write
    \begin{eqnarray}
    B_{ \omega, \ell} =B^*_{E, \ell} e^{-i \phi'_{\ell}(E)(\omega +E)} , \\
\overrightarrow{A}_{ \omega, \ell} =  \overleftarrow{A}^*_{E, \ell} e^{-i \chi'_{\ell}(E)(\omega +E)} , \label{aap}
\end{eqnarray}
    where $\chi_{\ell}(\omega) = \arg \overleftarrow{A}_{ \omega, \ell} $
    For $E \sigma >> 1$, the contribution from the $\overleftarrow{R}_{\omega, \ell} (r)$ solutions is negligible. We also approximate $e^{\frac{2\pi (E\pm \frac{1}{2}\xi)}{\kappa}}$ with $e^{\frac{2\pi E}{\kappa}}$. We obtain the following results.

\medskip

\noindent {\bf Case I:} both detectors are far from the horizon.   Eq. (\ref{main4}) becomes

\begin{eqnarray}
Z(E,  \Delta t) = \frac{1}{8 E^2r^2 r^{\prime 2}}  \frac{g_{\sigma}(\frac{1}{2} \Delta u)}{(  e^{ \frac{2\pi E}{\kappa}}-1)^2} T_E(\Theta)^2 \label{zedt2c}
\end{eqnarray}
 where
 \begin{eqnarray}
 T_E(\Theta) = \sum_{\ell = 0}^{\infty} (2 \ell +1) |B_{E, \ell}|^2 P_{\ell}(\cos \Theta).
 \end{eqnarray}

  \medskip

\noindent  {\bf Case II: } We find

\begin{eqnarray}
Z(E,  \Delta t) = \frac{1}{8 E^2r^2 r^{\prime 2}(  e^{ \frac{2\pi E}{\kappa}}-1)^2} \left\{ g_{\sigma}(\frac{1}{2} \Delta u) \left[\delta^2(\Theta)\right]^2 + g_{\sigma}(\frac{1}{2} \Delta v) S_E(\Theta)^2
\right. \nonumber \\
\left.
+ 2 \cos (2 E \Delta r_*) g_{\sigma}(\frac{1}{2} \Delta t) \delta^2(\Theta) S_E(\Theta) + C_E(\Theta, r+r'+\Delta t) + C_E(\Theta, r+r'-\Delta t)
\right\},
 \label{zedt3b}
\end{eqnarray}
where
\begin{eqnarray}
S_E(\Theta) = \sum_{\ell=0}^{\infty} (2 \ell + 1)  |\overrightarrow{A}_{E, \ell}|^2 P_{\ell}(\cos \Theta) = \delta^2(\Theta) - T_E(\Theta)\\
C_E(\Theta, x) = \sum_{\ell=0}^{\infty}\sum_{\ell'=0}^{\infty}(2 \ell + 1)(2 \ell' + 1) \overrightarrow{A}_{E, \ell}\overrightarrow{A}^*_{E, \ell'} P_{\ell}(\cos \Theta)P_{\ell'}(\cos \Theta)\nonumber \\
\times g_{\sigma}\left[\frac{1}{2} (\frac{1}{2} (\chi'_{\ell}(E) +\chi'_{\ell'}(E))+x\right].
\end{eqnarray}
In the derivation of Eq. (\ref{zedt3b}), we ignore rapidly oscillatory terms that are suppressed after coarse-graining of energy---see, Sec. 2.3.

The first two terms in Eq. (\ref{zedt3b})   are characterized by  $\Delta u = 0$ and $\Delta v = 0$, modulo the temporal accuracy $\sigma$ of the measurement. Incoming correlations ($\Delta v = 0$) are due to pairs of Hawking quanta that have been  reflected by the potential before they become detected.
The third term is non-zero  only for $\Delta t = 0$ irrespective of the distance between the two detectors.

The last two terms  corresponds to a pair of Hawking quanta, one detected while outgoing and the second detected while incoming after reflection on the potential. It involves interferences from different partial waves, because the effective time $\chi'_{\ell}(E)$ before a Hawking quantum is reflected depends on the angular momentum $\ell$.

\medskip

\noindent {\bf Case III:} one detector near and one far from the horizon. We assume that
 $r_* \rightarrow \infty$ and $r'_* \rightarrow -\infty$.

We obtain
\begin{eqnarray}
Z(E,  \Delta t) = \frac{1}{8 E^2r^2 r^{\prime 2}(  e^{ \frac{2\pi E}{\kappa}}-1)^2} \left[ D_E(\Theta, \Delta u) + F_E(\Theta, \Delta t - (r+r'))\right], \label{zedt4}
\end{eqnarray}
where
\begin{eqnarray}
D_E(\Theta, x) = \sum_{\ell=0}^{\infty}\sum_{\ell'=0}^{\infty}(2 \ell + 1)(2 \ell' + 1) B_{E, \ell}B^*_{E, \ell'}P_{\ell}(\cos \Theta)P_{\ell'}(\cos \Theta) \nonumber \\
\times g_{\sigma}\left[\frac{1}{2} (\frac{1}{2} (\phi'_{\ell}(E) +\phi'_{\ell'}(E))+x) \right],
\\
F_E(\Theta, x) = \sum_{\ell=0}^{\infty}\sum_{\ell'=0}^{\infty}(2 \ell + 1)(2 \ell' + 1) B_{E, \ell}\overrightarrow{A}^*_{E, \ell}B^*_{E, \ell'} \overrightarrow{A}_{E, \ell'}P_{\ell}(\cos \Theta)P_{\ell'}(\cos \Theta)\nonumber \\ \times g_{\sigma}\left[\frac{1}{2} (\frac{1}{2} (\phi'_{\ell}(E) - \chi'_{\ell}(E) +\phi'_{\ell'}(E) - \chi'_{\ell'}(E))+x)\right].
\end{eqnarray}

The first term in Eq. (\ref{zedt4}) corresponds to the detection of one outgoing  Hawking quantum near the horizon and  one quantum  that escaped the potential well. It involves interferences from different partial waves, because the effective time $\phi'_{\ell}(E)$ of transmission  depends on the angular momentum $\ell$. The second term corresponds to the correlations of one incoming quantum near the horizon after it has been reflected by the potential and of one quantum that has escaped.

\end{appendix}

 \end{document}